\documentclass[prl,twocolumn,showpacs,floatfix]{revtex4-1}

\usepackage[cp1250]{inputenc} 

\usepackage[T1]{fontenc}  

\usepackage{times}

\usepackage{graphicx}
  \graphicspath{{./Eps/}}
\usepackage{amsmath}
\usepackage{amssymb}

\begin{document}

\title{Bloch-Wannier theory of persistent current in a ring made of the band insulator: Exact result for one-dimensional lattice and estimates for realistic lattices}

\author{A. \surname{Mo\v{s}kov\'a}}

\author{M. \surname{Mo\v{s}ko}}
\email{martin.mosko@savba.sk}

\author{J. \surname{T\'obik}}

\affiliation{Institute of Electrical Engineering, Slovak
Academy of Sciences, 841 04 Bratislava, Slovakia}

\date{\today}

\begin{abstract}
It is known  that a mesoscopic normal-metal ring pierced by magnetic flux supports persistent
current. It was proposed recently that the persistent current exists also in a ring made
of a band insulator, where it is carried by electrons in the fully occupied valence band.
In this work, persistent currents in rings made of band insulators are analyzed theoretically. We first formulate a recipe which
determines the Bloch states of a one-dimensional (1D) ring from the Bloch states of an
infinite 1D crystal created by the periodic repetition of the ring. Using the recipe, we derive an expression
for the persistent current in a 1D ring made of an insulator with an arbitrary valence band $\epsilon(k)$.
To find an exact result for a specific insulator, we consider a 1D ring represented by a periodic lattice of $N$ identical sites with a single value of the on-site energy. If the Bloch states in the ring are expanded
over a complete set of $N$ on-site Wannier functions, the discrete on-site energy splits into the
energy band. At full filling, the band emulates the valence band of the band insulator and the ring is insulating. It carries
a persistent current equal to the product of $N$ and the derivative of the on-site energy with respect to the magnetic flux.
This current is not zero if one takes into account that the on-site Wannier function and consequently the on-site energy of each ring site depend on magnetic flux.
To derive the current analytically, we expand all $N$ Wannier functions of the ring over the infinite basis
of Wannier functions of the constituting infinite 1D crystal and eventually determine the crystal Wannier
functions by a method of localized atomic orbitals.
In terms of the crystal Wannier functions, the current at full
filling arises because the electron in the ring with $N$ lattice sites is allowed to make a single hop from site $i$ to its periodic replicas $i \pm N$.
In terms of the ring Wannier functions, the same current is due to the flux dependence of the Wannier functions basis, and
the longest allowed hop is from $i$ to $i \pm {\rm Int} (N/2)$ only.
Finally, we estimate the persistent current at full filling
in a few realistic systems. In particular, the rings made of real band insulators (GaAs, Ge, InAs)
with a fully filled valence band and empty conduction band are studied. The current decays with the ring length exponentially due to the exponential decay of the Wannier functions. In spite of that, it can be of measurable size.
\end{abstract}


\maketitle


\section{I. Introduction}

It is known that a conducting ring pierced by a constant magnetic flux can support persistent
electron current circulating around the ring \cite{Imry-book}. The persistent current
is well known to exist in a superconducting ring, and also in a normal-metal ring as long as it is
mesoscopic (of size comparable or smaller than the electron coherence length). Early work dealing
with persistent current and flux quantization in superconducting rings \cite{Byers} also contains
results relevant to the normal metal rings. Later it was mentioned \cite{Bloch} that one can have
circulating currents for free electrons in ballistic normal-metal rings, and the reference
\cite{Buttiker} proposed that persistent currents should exist in the disordered normal-metal rings.
The persistent currents were indeed observed in the disordered normal-metal rings
\cite{Levy,Chandrasekhar,Jariwala,Bluhm,Bleszynski} as well as in ballistic conducting rings
\cite{Mailly,Rabaud}. Details are reviewed in various papers \cite{Eckern,Saminadayar,Feilhauer2}.

One of us recently proposed \cite{Mosko} that persistent current exists also in a ring made of
a band insulator, where it is carried exclusively by Bloch electrons in the fully occupied
valence band. The effect is interesting for two reasons \cite{Mosko}. First, it seems to
be the only effect manifesting electron transport in a fully occupied valence band.
In a standard conductance measurement, the band insulator is biased by two
metallic electrodes and one observes the tunneling of conduction electrons from metal to
metal through the energy gap of the insulator. One does not observe the transport of valence band electrons in
the insulator. Second, in the fully occupied valence band, Bloch electrons cannot be scattered by
phonons or other electrons. Therefore, the electron coherence
length should be huge even at room temperature if the valence band is separated from the conduction
band by a large energy gap. The persistent current in the insulating ring should thus be observable
up to high temperatures as a temperature-independent effect as long as excitations into the
(empty) conduction band are negligible. By contrast, persistent currents in metallic rings
are observable only at low temperatures.
\cite{Bleszynski}.

We report on a study of persistent currents in rings made of band insulators using
the formalism of the Bloch and Wannier functions. It was partly motivated by a pioneering
analysis \cite{Cheung} of persistent currents in
a 1D ring-shaped periodic lattice composed of the lattice sites with a single value of the on-site energy.
The authors derived persistent current $I$ for spinless electrons at zero temperature assuming
that they move along the lattice by means of the nearest-neighbor-site hopping. The result reads
\begin{equation}
\label{perzistentny vysledok_tight_binding odd} I = -\frac {4\pi \Gamma_1}{N
\phi_0} \frac{\sin{\frac\pi N  N_e}}{\sin{\frac\pi N }}
\sin{\left[\frac{2\pi}{N}\frac{\phi}{\phi_0}\right]}, \quad -\frac{\phi_{0}}{2} \leq \phi < \frac{\phi_{0}}{2}
 \ ,
\end{equation}
for odd $N_e$ and
\begin{equation}
\label{perzistentny vysledok_tight_binding even} I = -\frac {4\pi \Gamma_1}{N
\phi_0} \frac{\sin{\frac\pi N  N_e}}{\sin{\frac\pi N }}
\sin{\left[\frac{\pi}{N}\left(\frac{2\phi}{\phi_0}-1\right)\right]}, \quad 0 \leq \phi < \phi_{0}
 \ ,
\end{equation}
for even $N_e$, where $N_e$ is the electron number in the ring, $N$ is the number of the lattice sites,
$\phi$ is the magnetic flux piercing the ring, $\phi_0 \equiv h/e$ is the flux quantum, and $\Gamma_1$ is the hopping amplitude.
Equations
\eqref{perzistentny vysledok_tight_binding odd} and \eqref{perzistentny vysledok_tight_binding even} show a nonzero current for a
conductor ($N_e<N$). However, they show a zero current for an insulator ($N_e=N$), which contradicts our claim that persistent current
exists also in an insulating ring.

Therefore, one of our goals is to give exact proof that a ring made of a band insulator does support a non-zero
persistent current carried by the fully filled valence band. We also provide a simple tool allowing to estimate
persistent currents in rings made of realistic band insulators. We finally present a derivation of an exact persistent current expression for a specific model insulator. This paper is structured as follows.

In Sect. II, we formulate
a recipe which determines the Bloch states of a one-dimensional (1D) ring from the Bloch states
of an infinite 1D crystal created by the periodic repetition of the ring.
In Sect. III we use the recipe to derive an expression for the persistent current in a 1D ring made of a band insulator with an arbitrary
valence band $\epsilon(k)$. The expression generally shows that the current is not zero,
although it does turn to zero for a cosine-shaped energy band, which was used in the analysis with nearest-neighbor-site hopping \cite{Cheung}.

To derive an exact result for a specific insulator, in Sect. IV we consider a 1D ring
represented by a periodic lattice of $N$ identical sites with a single value of the on-site energy. If the Bloch states in the ring are expanded
over a complete set of $N$ on-site Wannier functions, the discrete on-site energy splits into the
energy band. At full filling, the band emulates the valence band of the band insulator and the ring is insulating. We find that it carries
a persistent current equal to the product of $N$ and the derivative of the on-site energy with respect to the magnetic flux. An alternative
explanation of why Eqs. \eqref{perzistentny vysledok_tight_binding odd} and \eqref{perzistentny vysledok_tight_binding even} give zero current at full filling is that they do not take into account the flux-dependence of the on-site energy.

To get closer to an exact formula, in Sect. IV we expand all $N$ Wannier functions of the ring over the basis
of Wannier functions of the constituting infinite 1D crystal.
In terms of the crystal Wannier functions, the current at full
filling arises because the electron in the ring with $N$ lattice sites is allowed to make a single hop from site $i$ to its periodic replicas $i \pm N$.
Alternatively, in terms of the ring Wannier functions the same current is due to the flux dependence of the Wannier functions basis, and
the longest allowed hop is from $i$ to $i \pm {\rm Int} (N/2)$ only.

In Sect. V we eventually derive the crystal Wannier functions by a method of localized atomic orbitals (LCAO)
and express the persistent current at full filling by means of an exact formula.
The formula is found to be in good agreement with a numerical LCAO calculation.
The numerical data are provided for a 1D ring made of an artificial band insulator,
namely for a GaAs ring subjected to a periodic potential emulating the insulating lattice.

In Sect. VI we
estimate persistent currents in the rings made of real band insulators
(GaAs, Ge, InAs) with a fully filled valence band and empty conduction band. The current at full filling decays with the ring length exponentially due to the exponential decay of the Wannier function tails. In spite of that, it can be of measurable size.

Our results are summarized in Sect. VII. We relate them briefly to the paper by Kohn \cite{Kohn1},
which defines the insulating state as a property of the many-body ground state, and which contains a remark implicitly admitting
the existence of nonzero persistent current in insulating rings.

We note that some of the results of Sect. V were obtained in the conference contribution
\cite{Moskova} by means of a formally different approach. In another conference paper
\cite{Nemeth-1} the persistent current at full filling was analyzed for a 1D ring with
Kronig-Penney potential.

\section{II. Elementary theory and the general recipe}

To review basic concepts, we consider the circular 1D ring with circumference $L$, pierced by
magnetic flux $\phi$. The electrons in such ring are described by the Schr\" odinger equation
\begin{equation}
\left[ \frac{1}{2m}(-i\hbar\frac{d}{d x}+\frac{e\phi}{L})^{2}+V(x)
\right]\psi_{\phi}(x)=\varepsilon(\phi)\psi_{\phi}(x),
\label{eq:schr}
\end{equation}
where $x$ is the electron position along the ring circumference, $m$ is the electron mass, $V(x)$
is an arbitrary potential applied along the ring, $\varepsilon(\phi)$ is the electron
eigen-energy, and $\psi_{\phi}(x)$ is the electron wave function. Due to the ring geometry,
$\psi_{\phi}(x)$ has to fulfill the periodic condition
\begin{equation}
\psi_{\phi}(x) = \psi_{\phi}(x+L). \label{eq:period}
\end{equation}
We define the wave function $\varphi_{\phi}(x)$ by transformation
\begin{equation}
\varphi_{\phi}(x)=\exp\left(i\frac{2\pi}{L}\frac{\phi}{\phi_{0}}x\right)\psi_{\phi}(x).
\label{eq:sub}
\end{equation}
Setting $\psi_{\phi}(x)=\exp(-i\frac{2\pi}{L}\frac{\phi}{\phi_{0}}x)\varphi_{\phi}(x)$ into
(\ref{eq:schr}) and (\ref{eq:period}) we obtain the Schr\" odinger equation
\begin{equation}
\left[ -\frac{\hbar^2}{2m}\frac{d^2}{d x^2}+V(x)
\right]\varphi_{\phi}(x)=\varepsilon(\phi)\varphi_{\phi}(x)
\label{eq:schrsub}
\end{equation}
with the boundary condition
\begin{equation}
\varphi_{\phi}(x+L)=\exp(i2\pi\frac{\phi}{\phi_{0}})\varphi_{\phi}(x).
\label{eq:period1}
\end{equation}
Here the magnetic flux enters only the boundary condition.

In the ring geometry, $V(x)$ obeys the periodic condition
\begin{equation}
V(x) = V(x+L), \label{eq:Vperiod}
\end{equation}
which is the same as for the infinite 1D crystal with lattice constant $L$.
Therefore, equation \eqref{eq:schrsub} with condition \eqref{eq:Vperiod} is mathematically identical with the Schr\"
odinger equation
\begin{equation}
\left[ -\frac{\hbar^2}{2m}\frac{d^2}{d x^2}+V(x)
\right]\varphi_{k}(x)=\varepsilon(k) \varphi_{k}(x),
\label{eq:schrsub1}
\end{equation}
where $V(x)$ is the infinite periodic potential with period $L$, obtained by the periodic repetition of the ring potential $V(x)$.
Equation \eqref{eq:schrsub1} has the well known Bloch solution
\begin{equation}
\varphi_{k}(x)=\exp(ikx)u_{k}(x), \label{Blochwavefunction}
\end{equation}
where the function $u_{k}(x)$ fulfills the periodic condition
\begin{equation}
u_{k}(x)=u_{k}(x+L) \label{periodicpart},
\end{equation}
and $k$ is the electron wave vector from the interval $(-\infty, \infty)$.
Clearly, the wave function \eqref{eq:sub} and Bloch solution \eqref{Blochwavefunction} coincide for
$k = \frac{2 \pi}{L}\frac{\phi}{\phi_{0}}$. Let us discuss this coincidence in detail.

Consider first the ring with zero magnetic flux. To obtain the wave functions in such ring,
it suffices to take the Bloch function \eqref{Blochwavefunction} and to restrict it by the periodic condition
\begin{equation}
\varphi_{k}(x)=\varphi_{k}(x+L),  \label{periodicBloch}
\end{equation}
which is the condition \eqref{eq:period1} for $\phi = 0$.
Due to the condition  \eqref{periodicBloch}, the wave vector $k$ becomes discrete:
\begin{equation}
k=\frac{2 \pi}{L}n, \quad n=0,\pm 1,\pm 2,\ldots \, \label{eq:kdiskret}
\end{equation}
Thus, in the ring with zero magnetic flux and specified potential $V(x)$, the eigen-function $\varphi_{n}(x)$
and eigen-energy $\varepsilon_{n}$ can be calculated simply by setting $k=\frac{2 \pi}{L}n$ into the Bloch solutions
$\varphi_{k}(x)$ a $\varepsilon(k)$, calculated for the same potential $V(x)$ repeated with period $L$ from $x = -\infty$ to
$x = \infty$. This recipe can be generalized to nonzero magnetic flux as follows.

Arbitrary magnetic flux $\phi$ can be written in the form
\begin{equation}
\phi=n\phi_{0}+\phi^{,}, \label{eq:tokredukovany}
\end{equation}
where $\phi^{,}$ is the reduced flux from the range
$<-\frac{\phi_{0}}{2},\frac{\phi_{0}}{2})$ or alternatively from $<0,\phi_{0})$, and $n$ is one of the values
 $n=0,\pm 1,\pm 2,\ldots$. Setting \eqref{eq:tokredukovany}
into (\ref{eq:sub}) one can write (\ref{eq:sub}) in the form
\begin{equation}
\varphi_{n,\phi^{,}}(x)=\exp\left(i\frac{2\pi}{L}\frac{\phi^{,}}{\phi_{0}}x\right)\psi_{n,\phi^{,}}^{,}(x)
, \label{eq:sub12}
\end{equation}
where the function $\psi_{n,\phi^{,}}^{,}(x) \equiv \exp\left(i\frac{2\pi}{L}nx\right)
\psi_{\phi}(x)$ obeys the periodic condition $\psi_{n,\phi^{,}}^{,}(x)=\psi_{n,\phi^{,}}^{,}(x+L)$.
The boundary condition  \eqref{eq:period1} now reads
\begin{equation}
\varphi_{n,\phi^{,}}^{,}(x+L)=\exp(i2\pi\frac{\phi^{,}}{\phi_{0}})\varphi_{n,\phi^{,}}^{,}(x).
\label{eq:period1-2}
\end{equation}

Similarly, in the Bloch function theory it is customary to express the wave vector $k$
by means of the relation
\begin{equation}
k=\frac{2 \pi}{L}n+k^{'}, \label{eq:k redukovany}
\end{equation}
where  $k^{'}$ is the reduced wave vector from the first
Brillouin zone and the integer $n$ plays the role of the energy band number. Using
\eqref{eq:k redukovany} we can write the Bloch function \eqref{Blochwavefunction} as
\begin{equation}
\varphi_{n,k^{,}}(x)=\exp\left(ik^{,}x\right)u_{n,k^{,}}^{,}(x) ,
\label{eq:sub123}
\end{equation}
where the function $u_{n,k^{'}}^{'}(x) \equiv \exp\left(i\frac{2
\pi}{L}nx\right) u_{k}(x)$ fulfills the periodic condition
$u_{n,k^{'}}^{'}(x)=u_{n,k^{'}}^{'}(x+L)$. It is then easy to verify that
\begin{equation}
\varphi_{n,k^{,}}^{,}(x+L)=\exp(ik^{,}L) \varphi_{n,k^{,}}^{,}(x).
\label{eq:Blochperiod1-2}
\end{equation}

The equations \eqref{eq:sub12} and \eqref{eq:period1-2}  coincide with the equations
\eqref{eq:sub123} and \eqref{eq:Blochperiod1-2}, respectively, if
\begin{equation}
k^{'} = \frac{2 \pi}{L}\frac{\phi^{,}}{\phi_{0}}, \label{eq:2
redukovanydoBrzony}
\end{equation}
or alternatively, if
\begin{equation}
k = \frac{2 \pi}{L}(n + \frac{\phi^{,}}{\phi_{0}}). \label{eq:2 n
kneredukovany}
\end{equation}
One can thus formulate the following general recipe. In the ring with a known potential $V(x)$, the
eigen-function $\varphi_{n,\phi^{,}}^{,}(x) \equiv \varphi_{\phi}(x)$ and eigen-energy
$\varepsilon_{n}(\phi^{'}) \equiv \varepsilon(\phi)$ can be calculated by setting \eqref{eq:2
redukovanydoBrzony} or \eqref{eq:2 n kneredukovany} into the Bloch solutions $\varphi_{n,k^{,}}(x)
\equiv \varphi_{k}(x)$ a $\varepsilon_{n}(k^{,}) \equiv \varepsilon(k)$, calculated for the same
potential $V(x)$ repeated with period $L$ from $x = -\infty$ to $x = \infty$.

\begin{figure}[t]
\centerline{\includegraphics[clip,width=0.90\columnwidth]{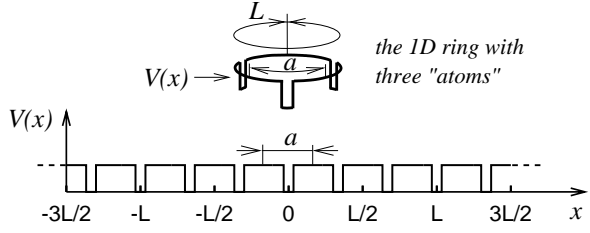}} \vspace{-0.15cm} \caption{The
top figure shows schematically the 1D ring composed of the periodic lattice with $N$ lattice sites ($N=3$), periodicity $a$, and length
$L = Na$.  For simplicity, the lattice potential of the ring is modeled by a square-well potential $V(x)$. The bottom figure depicts the
infinite 1D periodic potential obtained by the periodic repetition of the ring potential. }
 \label{Fig:N-atomicring}
\end{figure}

The general recipe holds for any
potential $V(x)$ obeying the cyclic condition \eqref{eq:Vperiod}. It therefore holds also
for any potential which additionally obeys the periodic condition
\begin{equation}
V(x) = V(x+a), \label{eq:Vperiod-a}
\end{equation}
where $a=L/N$ and $N$ is the (integer) number of periods $a$ within the period $L$. This type of potential will be considered in the rest of the paper.
Figure \ref{Fig:N-atomicring} illustrates how
an infinite 1D crystal is created from a 1D ring with lattice period $a$ and length $L = Na$.
Such crystal is still described by Schr\"
odinger equation \eqref{eq:schrsub1}, except that now $V(x) = V(x+a)$ and consequently $u_k(x)=u_k(x+a)$. However, the condition $u_k(x)=u_k(x+L)$
holds as well because $V(x) = V(x+L)$. We can rederive the general recipe briefly, if we compare $u_k(x)=u_k(x+L)$ with $\psi_{\phi}(x) = \psi_{\phi}(x+L)$ and
$\varphi_k(x) = \exp(ikx) u_k(x)$ with $\varphi_{\phi}(x)=\exp\left(i\frac{2\pi}{L}\frac{\phi}{\phi_{0}}x\right)\psi_{\phi}(x)$.
We see that $\varphi_{\phi}(x)$ coincides with $\varphi_k(x)$ for $k=\frac{2\pi}{L}\frac{\phi}{\phi_{0}}$.

It remains to express the persistent current. The Bloch electron in state
$(n, k^{'})$ moves with velocity
$v_{n}(k^{'})=\frac{1}{\hbar}\frac{\partial{\varepsilon_{n}(k^{'})}}{\partial
{k^{'}}}$. We set into the Bloch electron velocity the formula \eqref{eq:2
redukovanydoBrzony}. We obtain the expression $v_{n}(\phi^{'})=\frac{L}{e}\frac{\partial{\varepsilon_{n}(\phi^{'})}}{\partial{\phi^{'}}}$,
which is the electron velocity in state $(n,
\phi^{'})$ in the ring. The current carried by the electron in state $(n, \phi^{'})$
reads
\begin{equation}
I_{n}(\phi^{'})=-\frac{ev_{n}(\phi^{'})}{L}=-\frac{\partial{\varepsilon_{n}(\phi^{'})}}{\partial{\phi^{'}}}.
\label{eq:prud}
\end{equation}
The total persistent current circulating in the ring is
\begin{equation}
I(\phi^{'})=-\frac{\partial}{\partial{\phi^{'}}}\sum_{n}\varepsilon_{n}(\phi^{'}),
\label{eq:prud1}
\end{equation}
where we sum (at zero temperature) over all occupied states $n=0,\pm 1,\pm 2,\ldots$ up to the Fermi level.
In the following text we use the formulas \eqref{eq:prud} and \eqref{eq:prud1} with symbol $\phi^{'}$
changed to $\phi$, where $\phi \in <-\frac{\phi_{0}}{2},\frac{\phi_{0}}{2})$ or alternatively $\phi \in <0,\phi_{0})$.

\section{III. Persistent current in 1D ring made of a crystal with arbitrary valence band $\varepsilon(k)$}
Consider first an infinite 1D crystal created from a 1D ring with the lattice period $a$ and length $L = Na$ as it is illustrated in
Fig. \ref{Fig:N-atomicring}. Let $\varepsilon_{v}(k)$ be the energy dispersion of the valence band of Bloch electrons in that crystal. The wave vector $k$ in $\varepsilon_{v}(k)$ is supposed to be from the Brillouin zone $<-\frac{\pi}{a},\frac{\pi}{a}>$. In what follows we skip the valence band index ($v$) for simplicity and we use notation $\varepsilon(k)$. We can express $\varepsilon(k)$ by means of the infinite Fourier expansion
\begin{eqnarray}
\varepsilon(k)&=&a_0+a_1\cos(ka)+
\nonumber
\\
&&
a_2\cos(2ka)+\dots a_N\cos(Nka)+\dots,
\nonumber
\\
\label{oneparticleenergyexpansion}
\end{eqnarray}
where
\begin{equation}
a_0 =
\frac {1}{\pi/a}\int_{0}^{\pi/a}\varepsilon(k)dk
\label{a_0}
\end{equation}
and
\begin{equation}
a_j =
\frac {2}{\pi/a}\int_{0}^{\pi/a}\varepsilon(k)\cos(jka)dk , \quad j = 1, 2, \dots
\label{a_j}
\end{equation}

Now we apply the general recipe from Sect. II.
We set for $k$ in Eq. \eqref{oneparticleenergyexpansion} the equation
$k = \frac{2 \pi}{Na}(n + \frac{\phi}{\phi_{0}})$. We obtain the expression for the ring energies $\varepsilon_{n}(\phi)$:
\begin{equation}
\varepsilon_{n}(\phi) \equiv \varepsilon(k_n(\phi))= \sum_{j=0}^\infty a_j
\cos \left(k_n(\phi)\ ja\right)\ ,
\label{e_k_n text}
\end{equation}
where
\begin{equation}
k_n(\phi) \equiv \frac{2\pi}{Na}(n+\frac{\phi}{\phi_0}) \ .
\label{k_n text}
\end{equation}
As shown in appendix A, expansion \eqref{e_k_n text}
can also be written in the form
\begin{equation}
\varepsilon_{n}(\phi)=
\Omega_0 +
\sum_{j=1}^M
\Bigl\{
\Omega^{R}_j
\cos \left(k_n(\phi)\ ja\right)
+
\Omega^{I}_j
\sin \left(k_n(\phi)\ ja\right)
\Bigr \} \ ,
\label{e_k_n_finite}
\end{equation}
where $M = {\rm Int} (N/2)$ and the Fourier coefficients $\Omega$ are given by equations
\eqref{Omega odd} and \eqref{Omega even}. Expressions \eqref{e_k_n text} and \eqref{e_k_n_finite} are mathematically equivalent,
but expression \eqref{e_k_n_finite} reveals an important difference from the Bloch energy \eqref{oneparticleenergyexpansion}.
Generally, the Bloch energy \eqref{oneparticleenergyexpansion} consists of an infinite number of Fourier terms,
with the coefficients $a_j$ being independent on magnetic flux $\phi$. On the other hand, expression \eqref{e_k_n_finite} shows that
the Bloch energy in the ring can be expressed through a finite number of Fourier terms, with the Fourier coefficients $\Omega$
depending on $\phi$ [see Eqs. \eqref{Omega odd} and \eqref{Omega even}].

In the rest of this section we use Eq.
\eqref{e_k_n text}. The same results can be obtained by means of Eq. \eqref{e_k_n_finite}, but derivations are
more cumbersome.

Setting Eq. \eqref{e_k_n text} into Eq. \eqref{eq:prud1} we obtain the persistent current
\begin{eqnarray}
\nonumber
I(\phi)  &=&\frac{2\pi}{N}\ \frac{1}{\phi_0}\   \sum_{j=1}^{\infty} j\ a_j \\
\nonumber
&\times& \Biggl[
\cos\left(\frac{2\pi}{N}\frac{\phi}{\phi_0}\ j\right)\sum_{n}\sin\left(\frac{2\pi}{N}j\cdot
n\right)
\\
&+&\sin\left(\frac{2\pi}{N}\frac{\phi}{\phi_0}j\right) \sum_{n}\cos\left(\frac{2\pi}{N}j\cdot
n\right) \Biggr] \  . \label{currentVSflux2}
\end{eqnarray}

If the ring contains $N_e$ spinless electrons and $N_e$ is odd,
summation over $n$ in Eq. \eqref{currentVSflux2} includes the occupied states
$n=0,\pm1,\pm2,\dots,\pm(N_e-1)/2$. After simple manipulations
\begin{eqnarray}
\nonumber
I(\phi)  &=& \frac{2\pi}{N}\ \frac{1}{\phi_0}\  \sum_{j=1}^{\infty} j\ a_j
\sin\left(\frac{2\pi}{N}\frac{\phi}{\phi_0}\ j\right)\\
&\times& \left[2
\sum_{n=0}^{(N_e-1)/2}\cos\left(\frac{2\pi}{N}j\cdot n\right)-1 \right] \  . \label{currentVSflux3a}
\end{eqnarray}
Performing the summation over $n$
we arrive to the formula
\begin{eqnarray}
\nonumber
I(\phi)  &=&\frac{2\pi}{N}\ \frac{1}{\phi_0} \Biggl[ \hspace{-6mm}\sum_{\substack{j=1 \\
\qquad j\neq N,2N,\dots}}^{\infty} \hspace{-4mm}
j  a_j\sin\left(\frac{2\pi}{N}\frac{\phi}{\phi_0}\
j\right)\frac{\sin\left(\frac{\pi}{N}jN_e\right)}{\sin\left(\frac{\pi}{N}j\right)} \\
&+& N_e  \hspace{-3mm} \sum_{j=N,2N,\dots}^{\infty} \hspace{-2mm} ja_j\sin\left(\frac{2\pi}{N}\frac{\phi}{\phi_0}j\right) \Biggr]
\ , \label{currentVSflux4}
\end{eqnarray}
where  $\phi \in <-\frac{\phi_{0}}{2},\frac{\phi_{0}}{2})$. Note that the first sum in
\eqref{currentVSflux4}
omits the terms $j = N, 2N, 3N, \dots$, included in the second sum.

If $N_e$ is even and $\phi \in <0,\phi_{0})$,
we sum in Eq. \eqref{currentVSflux2} over the occupied states
$n=0,\pm1,\pm2,\dots,\pm(N_e/2-1),-N_e/2$. Similar manipulations as before lead to the result
\begin{eqnarray}
\nonumber
I(\phi)  &=&\frac{2\pi}{N}\ \frac{1}{\phi_0} \\
\nonumber
&\times& \Biggl[ \hspace{-6mm}\sum_{\substack{j=1 \\
\qquad j\neq N,2N,\dots}}^{\infty} \hspace{-4mm}
j  a_j\sin\left(\frac{\pi}{N}\left(\frac{2\phi}{\phi_0}-1\right)\
j\right)\frac{\sin\left(\frac{\pi}{N}jN_e\right)}{\sin\left(\frac{\pi}{N}j\right)} \\
&+& N_e  \hspace{-3mm} \sum_{j=N,2N,\dots}^{\infty} \hspace{-2mm}
ja_j\sin\left(\frac{2\pi}{N}\left(\frac{\phi}{\phi_0}\right)j\right) \Biggr] \ ,
\label{currentVSflux4 even}
\end{eqnarray}
where $\phi \in <0,\phi_{0})$. The results \eqref{currentVSflux4} and \eqref{currentVSflux4 even}
hold for the conducting ($N_e < N$) as well as insulating ($N_e = N$) rings.

For $N_e=N$ the first sum on the right hand side of Eqs. \eqref{currentVSflux4} and \eqref{currentVSflux4 even}
becomes zero and both equations take the form
\begin{equation}
I(\phi)=-(2\pi/\phi_0) \hspace{-3mm} \sum_{j=N,2N,\dots}^{\infty}
\hspace{-2mm} j  a_j\sin\left(\frac{2\pi}{N}\frac{\phi}{\phi_0}j\right), \label{currentVSfluxarb}
\end{equation}
where $\phi \in <-\frac{\phi_{0}}{2},\frac{\phi_{0}}{2})$ and $N$ can be even as well as odd.
Equation \eqref{currentVSfluxarb} expresses the persistent current in a 1D ring made of a band insulator with
an arbitrary valence band $\varepsilon_{k}$.

An important property of the formula \eqref{currentVSfluxarb} can be recognized without specifying
the dispersion law $\varepsilon_{k}$. The dispersion $\varepsilon(k)$ is a periodic function of period $2 \pi/a$. As long as it is analytic in the whole Brillouin zone $<-\frac{\pi}{a},\frac{\pi}{a}>$, the Fourier coefficients $a_j$ in the expansion \eqref{oneparticleenergyexpansion} obey the inequality \cite{Pinsky}
\begin{equation}
|a_j| \leq C e^{- \xi j}, \label{Fouriercoefprop}
\end{equation}
where $C$ and $\xi$ are positive constants. Equation \eqref{currentVSfluxarb} combined with inequality \eqref{Fouriercoefprop}
shows
that the persistent current in the insulating ring is in general not zero albeit it decreases with the ring length
at least exponentially.
 In the next sections, we will demonstrate for various specific insulating rings the exponential decrease, and we will see that the current can be of measurable size in realistic systems.

The pioneer results \eqref{perzistentny vysledok_tight_binding
odd} and \eqref{perzistentny vysledok_tight_binding even} can be obtained from Eqs.
\eqref{currentVSflux4} and \eqref{currentVSflux4 even} if we keep only the
term $j=1$ and we put $a_1 \equiv -2\Gamma_1$. Equations \eqref{perzistentny vysledok_tight_binding
odd} and \eqref{perzistentny vysledok_tight_binding even} predict for insulating rings ($N_e=N$) a zero persistent current, which contradicts
equation \eqref{currentVSfluxarb}. Origin of this
contradiction is evident. Keeping only the term $j=1$ means that only the nearest-neighbor-site hopping is operative.
This approach, also known as tight-binding approximation, simplifies the general dispersion law \eqref{oneparticleenergyexpansion}
to the cosine shaped dispersion $\varepsilon(k) \propto \cos(ka)$.
However, equation \eqref{currentVSfluxarb} shows that nonzero persistent current in the insulating ring of length $N$
is (mainly) due to the Fourier coefficient $a_N$ which is in general not zero for a realistic dispersion law.
We will see in the next section that $a_N$ physically represents the hopping amplitude for a single hop from site $i$ to its periodic replicas $i \pm N$.

\section{IV. Persistent current in insulating 1D ring in the Wannier functions formalism}

We consider again the 1D ring with lattice period $a$ and length $L = Na$, sketched in Fig. \ref{Fig:N-atomicring}.
Simultaneously, we consider the infinite 1D crystal obtained by the periodic repetition of the 1D ring, as it is illustrated in Fig. \ref{Fig:N-atomicring}. For the sake of clarity we recall a few major points. First, the Bloch function $\varphi_{k}(x)$ and Bloch energy $\varepsilon(k)$ in the infinite crystal
are described by Schr\" odinger equation
\begin{equation}
H \varphi_{k}(x)=\varepsilon(k)\varphi_{k}(x) \ ,
\nonumber
\end{equation}
\begin{equation}
H = -\frac{\hbar^2}{2m}\frac{d^2}{d x^2}+V(x), \quad V(x) = V(x+a).
\label{eq:schrsub1 wannier}
\end{equation}
Second, according to the general recipe of Sect. II, the Bloch function and Bloch energy in the ring, $\varphi_{n,\phi}(x)$
and $\varepsilon_n(\phi)$, are given by equations
\begin{equation}
\varphi_{n,\phi}(x) = \varphi_{k = k_n(\phi)}(x), \quad \varepsilon_n(\phi) = \varepsilon(k= k_n(\phi)) \ ,
\nonumber
\end{equation}
\begin{equation}
k_n(\phi) \equiv \frac{2\pi}{Na}(n+\frac{\phi}{\phi_0}) \ .
\label{eq:Bloch states ring wannier}
\end{equation}
Third, the Bloch vector $k$
is assumed to be from the Brillioun zone $<-\frac{\pi}{a},\frac{\pi}{a}>$, or alternatively from $<0,\frac{2\pi}{a}>$. This means that
$\varepsilon(k)$ is the dispersion of a single energy band and $\varphi_{k}(x)$ are the Bloch functions of that band. As before, the functions $\varphi_{k}(x)$ and $\varepsilon(k)$ do not involve any band index, and $\varepsilon(k)$
is supposed to represent the valence band of the crystal. Below we define the Wannier functions by restricting us to the Bloch states of that valence band.

\subsection{A. Wannier functions of the infinite crystal}

We first define the Wannier functions for the crystal. According to a standard textbook definition,
the Wannier function $W(x-la)$ at the lattice site $x=la$ is defined as
\begin{equation}
W(x-la)\equiv W_l(x)=\frac{a}{2\pi}\int_{-\frac{\pi}{a}}^{\frac{\pi}{a}}\
e^{-ikla}\Phi_k(x)dk \ ,
\label{wanierCrystal}
\end{equation}
where $\Phi_k(x)$ is the Bloch function of the infinite crystal. We note that $\Phi_k(x)$ is the same Bloch function
as $\varphi_k(x)$ in equation \eqref{eq:schrsub1 wannier}. The only difference is that $\Phi_k(x)$ is normalized as
\begin{equation}
\int_{0}^{a}\ \mid\Phi_k(x)\mid ^2 \ dx = 1
\
\label{blochCrystalNorm1}
\end{equation}
while the normalization of $\varphi_k(x)$ in the infinite crystal is unspecified.
In the following text, $\Phi_k(x)$ represents the normalized Bloch function of the infinite crystal
and the symbol $\varphi_k(x)$ with $k=k_n(\phi)$ represents the Bloch function of the ring.
[In the ring $\varphi_k(x)$ is normalized by equation \eqref{blochRingNorm}.]

Since the crystal is infinite, $k$ is a continuous variable and the Bloch functions fulfill
the orthogonality condition
\begin{equation}
\int_{-\infty}^{\infty}\Phi_{k^{\prime}}(x)^*\Phi_k(x) \ dx
= \frac{2\pi}{a}\delta(k^{\prime}-k) \ .
\label{blochCrystalNorm2}
\end{equation}
One can then easy verify the orthogonality condition for the Wannier functions,
\begin{equation}
\int_{-\infty}^{\infty}W_{l^{\prime}}^*(x)W_l(x)\ dx = \delta_{l,l^{\prime}} \ .
\label{wanierCrystalNorm}
\end{equation}
Finally, definition \eqref{wanierCrystal} can also be written in the inverted form
\begin{equation}
\Phi_k(x)=\sum_{l}\ e^{ikla}W_l(x)
\label{blochCrystalInverseT}
\end{equation}
where one sums over all lattice points of the (infinite) crystal.

We define the matrix elements $\Gamma_j$ by equations
\begin{equation}
\Gamma_{l',l} =
-\int_{-\infty}^{\infty} W_{l'}(x)\hat H W_l(x)dx = \Gamma_{l-l'} \ , \ \ \ \Gamma_j \equiv \Gamma_{l-l'},
\label{hoppingcrystal}
\end{equation}
where $l, l' = 0, \pm 1, \pm 2, \dots \pm \infty$ and it is taken into account that $\Gamma_{l',l}$ depends only on the difference $j = l-l'$.
The Bloch energy $\epsilon(k)$ can be written as $\epsilon(k) =  \langle\Phi_{k}\vert \hat H \vert
\Phi_{k}\rangle / \langle \Phi_{k}\vert\Phi_{k}\rangle$. If we set for $\Phi_{k}$ equation  \eqref{blochCrystalInverseT},
we easy find
\begin{equation}
\varepsilon(k)={\it \Gamma}_0 +2\sum_{j=1}^{\infty} {\it \Gamma}_j\ \cos{jka} .
\label{Blochenergy_wanier}
\end{equation}
Comparison of Eq. \eqref{Blochenergy_wanier} with Eq. \eqref{oneparticleenergyexpansion} shows that $a_j = 2 \Gamma_j$.
Obviously,
\begin{equation}
\Gamma_{j} = \frac {1}{\pi/2a}\int_{-\pi/a}^{\pi/a}\varepsilon(k)\cos(jka)dk ,
\label{hoppingcrystal 2}
\end{equation}
which is an alternative form of equation \eqref{hoppingcrystal}.

Finally, setting $a_j = 2 \Gamma_j$ into the result \eqref{currentVSfluxarb} we get
\begin{equation}
I=-(4\pi/\phi_0) \hspace{-3mm} \sum_{j=N,2N,\dots}^{\infty}
\hspace{-2mm} j  \Gamma_j\sin\left(\frac{2\pi}{N}\frac{\phi}{\phi_0}j\right) \ .
\label{persistentfullresult final}
\end{equation}
Equation \eqref{persistentfullresult final} together with Eq. \eqref{hoppingcrystal} express the persistent current in the insulating ring through the Wannier functions of the constituting infinite crystal. We defer discussion of these expressions to subsection C, where equation \eqref{persistentfullresult final} is derived from the Wannier functions of the ring.

\subsection{B. Wannier functions of the ring}

Let us define the Wannier functions of the ring.
In the ring with $N$ lattice sites and a fixed value of magnetic flux only the discrete Bloch vectors $k_n(\phi)$ are allowed, where $k_n(0) \in <-\frac{\pi}{a},\frac{\pi}{a}>$ or alternatively $k_n(0) \in <0,\frac{2\pi}{a}>$. In both cases there are $N$ allowed values
of $n$ (specified below), which means that the originally continuous band $\varepsilon(k)$
consists of $N$ discrete levels $\varepsilon_n(\phi)$. The eigen-functions of these levels
represent a complete set of $N$ Bloch functions obeying the orthogonality condition
\begin{equation}
\int_0^{Na}\ \varphi_{{k}^{\prime}}^*(x)\varphi_{k}(x)\ dx =
\delta_{{k},{k}^{\prime}}
=\delta_{{n},{n}^{\prime}}
\ ,
\label{blochRingNorm}
\end{equation}
where $k = k_n(\phi)$. Consequently, in the 1D ring with $N$ lattice sites the Wannier function at site $x=la$
can be defined as
\begin{equation}
w(x-la)\equiv
w_l(x)=
\frac 1{\sqrt{N}}\sum_{k} e^{-ikla}\varphi_{k}(x) \ ,
\label{wanierRing}
\end{equation}
where $k = k_n(\phi)$ and summation over $k$ means the summation over the $N$ allowed values of $n$.
Definition \eqref{wanierRing} coincides with definition (3.7) in Kohn's paper \cite{Kohn1}, except that Kohn
assumes $k = k_n(0)$.

In the following calculations we choose the Brillouin zone $k_n(0) \in <0,\frac{2\pi}{a}>$,
for which the allowed $n$ are simply
\begin{equation}
n=0, 1, \dots, N-1
\label{allowed latter}
\end{equation}
without regards to the parity of $N$ and sign of $\phi$.

Using Eqs. \eqref{blochRingNorm}  and \eqref{wanierRing} we can easy verify the orthogonality condition
\begin{equation}
\int_0^{Na}w_{l^\prime}(x)^*w_l(x)\ dx = \delta_{l^{\prime},l} \ .
\label{wanierRingNorm}
\end{equation}
Finally, definition \eqref{wanierRing} can also be written in the inverted form
\begin{equation}
\varphi_k(x)=\frac 1{\sqrt{N}}\sum_{l=0}^{N-1}e^{ikla}w_l(x) \ .
\label{blochRing}
\end{equation}
In this paper the Wannier functions $w_l(x)$ are often called the ring Wannier functions
while the Wannier functions $W_l(x)$ are called the crystal Wannier functions. The ring Wannier functions are magnetic-flux dependent,
the crystal Wannier functions are not.

We note that the results obtained in the rest of this section can also be obtained by choosing $k_n(0) \in <-\frac{\pi}{a},\frac{\pi}{a}>$. In that case the allowed values of $n$ are
\begin{equation}
n=0,\pm1,\pm2,\dots,\pm(N-1)/2
\label{allowed former odd}
\end{equation}
for odd $N$, while for even $N$ one has
\begin{equation}
n=0,\pm1,\pm2,\dots,\pm(N/2-1),-N/2 \quad \text{for $\phi > 0$}
\label{allowed former even pos}
\end{equation}
and
\begin{equation}
n=0,\pm1,\pm2,\dots,\pm(N/2-1),+N/2 \quad \text{for $\phi < 0$}.
\label{allowed former even neg}
\end{equation}
Evidently, it is more straightforward to work with Eq. \eqref{allowed latter}.

Now we express the persistent current at full filling in terms of the ring Wannier functions.
We write $\varepsilon_n(\phi)$ in the form
\begin{equation}
\varepsilon_n(\phi)=
\int_0^{Na}
 \varphi_{k_n(\phi)}^{\ast}(x)
 H \varphi_{k_n(\phi)}(x)dx \ .
\label{eigenenergy}
\end{equation}
We need to rewrite the right hand side of Eq. \eqref{eigenenergy} in terms of the matrix elements
\begin{equation}
T_{l^{\prime},l}=\int_0^{Na}\ w^*_{l^{\prime}}(x)\ H \ w_l(x) \ dx \ ,
\label{Tlprimel}
\end{equation}
where $l,l^{\prime}= 0, 1, \dots N-1$. We set into Eq. \eqref{Tlprimel} the equation \eqref{wanierRing}. We find
\begin{eqnarray}
T_{l^{\prime},l}
&=&
\frac 1N\sum_{{n}^{\prime}}\sum_{n} e^{i{k_n}^{\prime}(\phi)l^{\prime}a}e^{-i{k_n(\phi)}la}
\nonumber \\
&\times&
\int_0^{Na}\varphi^*_{{k_n}^{\prime}(\phi)}(x)H\varphi_{k_n(\phi)}(x)\ dx
\nonumber \\
&=&
\frac 1N e^{i\frac{2\pi}{N}\frac{\phi}{\phi_0}(l^{\prime}-l)}\sum_{n=0}^{N-1}
e^{i\frac{2\pi}{N}n(l^{\prime}-l)}\varepsilon_{n}(\phi) \ .
\nonumber \\
\label{Tproperties}
\end{eqnarray}
Since  $T_{l^{\prime},l}$ depends only on the difference between $l$ and $l^{\prime}$, we can write
\begin{equation}
T_{l^{\prime},l} = T_{l^{\prime}-l} \equiv T_j
= \frac 1N e^{i\frac{2\pi}{N}\frac{\phi}{\phi_0}j}\sum_{n=0}^{N-1}
e^{i\frac{2\pi}{N}nj}\varepsilon_{n}(\phi)\  .
\label{Tj}
\end{equation}
Inverting  Eq. \eqref{Tj} we get
\begin{equation}
\varepsilon_{n}(\phi)= \sum_{j=0}^{N-1}\ e^{-ik_n(\phi)ja}T_j \ .
\label{eigenenergy2}
\end{equation}
By means of Eq. \eqref{eigenenergy2}, the ground state energy of the fully filled valence band is readily obtained in the form
\begin{equation}
E(\phi) = \sum_{n=0}^{N-1}\varepsilon_n(\phi)=N\cdot T_0(\phi) \ .
\label{groundstateenergy}
\end{equation}
Finally, the persistent current at full-filling reads
\begin{equation}
I \ = \ -  \frac{dE}{d\phi} \ = \
- N \frac{dT_0}{d\phi}  \ .
\label{persistentfull}
\end{equation}
One can see from Eq. \eqref{Tlprimel} that the matrix element $T_0$
represents the on-site energy of the Wannier state $w_l(x)$.
Since  $w_l(x)$ depends on magnetic flux, so does $T_0$.
This however means that the current \eqref{persistentfull} is in general not zero. We have thus arrived at
the same conclusion as in the preceding section, where the persistent current in the insulating ring (Eq. \ref{currentVSfluxarb})
was derived from the Bloch states and general recipe.

\subsection{C. Ring Wannier functions in the basis of crystal Wannier functions}

The result \eqref{persistentfullresult final}, or equivalently the result \eqref{currentVSfluxarb},
can be obtained from the
result \eqref{persistentfull}, if the ring Wannier functions are expanded over the basis of the
crystal Wannier functions. To show this we proceed as follows.

The Bloch functions of the ring are given by Eq. \eqref{eq:Bloch states ring wannier} and normalized as
$\int_0^{Na}\ \varphi_{k}^*(x)\varphi_{k}(x)\ dx = 1$, where $k = k_n(\phi)$ [see Eq. \eqref{blochRingNorm}]. The Bloch functions
$\phi_k(x)$ coincide with $\varphi_{k}(x)$ except for the normalization constant. As they are normalized by means of Eq. \eqref{blochCrystalNorm1}, in the ring we have the relation
\begin{equation}
\varphi_k(x)=\frac 1{\sqrt{N}}\phi_k(x)\ , \quad \quad k=k_n(\phi) \ ,
\label{recipe}
\end{equation}
We set for $\phi_k(x)$ in Eq. \eqref{recipe} the definition \eqref{blochCrystalInverseT}. Then we set Eq. \eqref{recipe}
into the right hand side of Eq. \eqref{wanierRing}. We obtain
\begin{equation}
w_l(x)
=
\frac 1N \sum_k\sum_{l'=-\infty}^{\infty} e^{-ik(l-l')a}  W_{l'}(x) \ ,
\label{wanierringcrystal}
\end{equation}
where $k = k_n(\phi)$ and summation over $k$ means the summation over $n=0, 1, \dots, N-1$.
Using substitution $l'=Nr+s$, where $s=0,1, \dots (N-1)$
and $r=0,\pm 1, \pm 2, \dots \pm \infty$, we further obtain
\begin{eqnarray}
w_l(x)&=&
\sum_{r=-\infty}^{\infty} \sum_{s=0}^{N-1}
\ W_{Nr+s}(x)e^{i\frac{2\pi}{N}\frac{\phi}{\phi_0}(rN+s-l)}
\nonumber \\
&\times&
\frac 1N \sum_{n=0}^{N-1}e^{i\frac{2\pi}{N}n(s-l)} \ .
\nonumber \\
\label{wanierGeneral_1}
\end{eqnarray}
Eventually
\begin{equation}
w_l(x)
=
\sum_{r=-\infty}^{\infty} e^{i2\pi\frac{\phi}{\phi_0}r} \ W_{Nr+l}(x) \ .
\label{wanierGeneral}
\end{equation}
Equation \eqref{wanierGeneral} expresses the ring Wannier function at the ring site $l$, $w_l(x)$,
through the crystal Wannier functions $W_{l'}(x)$.
For $\phi = 0$ equation \eqref{wanierGeneral} coincides with equation (5.9) in Kohn's paper \cite{Kohn1}.

Setting Eq. \eqref{wanierGeneral} into Eq. \eqref{Tlprimel} we obtain
\begin{eqnarray}
&&
T_j = \int_0^{Na}\ w^*_{j}(x)\ \hat{H} \ w_0(x) \ dx
\nonumber \\
&=&
\hspace{-3mm}
\sum_{r=-\infty}^{\infty}\sum_{r^{\prime}=-\infty}^{\infty}
\hspace{-2mm}
e^{i2\pi\frac{\phi}{\phi_0}(r-r^{\prime})}
\hspace{-1.5mm}
\int_0^{Na}  \hspace{-2mm}   W_{Nr^{\prime}+j}(x) \hat{H}  W_{Nr}(x)dx
\nonumber \\
&=&
\hspace{-3mm}
\sum_{s=-\infty}^{\infty}\hspace{-1.5mm} e^{-i2\pi\frac{\phi}{\phi_0}s}
\hspace{-1.5mm}
\sum_{r^=-\infty}^{\infty}
\hspace{-0.1mm}
\int_{0-rNa}^{Na-rNa} \hspace{-2.5mm} \ W_{Ns+j}(x) \hat{H}  W_{0}(x)dx
\nonumber \\
&=&
\sum_{s=-\infty}^{\infty}e^{-i2\pi\frac{\phi}{\phi_0}s}
\int_{-\infty}^{\infty}\ W_{Ns+j}(x) \hat{H}  W_{0}(x)dx \ .
\nonumber \\
\label{Tgeneral1}
\end{eqnarray}
By means of the matrix elements $\Gamma_j$, equation \eqref{Tgeneral1} can be written as
\begin{equation}
T_j = \sum_{s=-\infty}^{\infty}e^{-2\pi i\frac{\phi}{\phi_0}s}\Gamma_{sN+j} \ .
\label{Tgeneral}
\end{equation}
For $j = 0$ equation \eqref{Tgeneral} gives
\begin{equation}
T_0 =\Gamma_0 + 2 \sum_{s=1}^{\infty} \
\Gamma_{sN}
\cos(2\pi \frac{\phi}{\phi_0}s) \ .
\label{T0general}
\end{equation}
We substitute Eq. \eqref{T0general} into the right hand side of Eq. \eqref{persistentfull}. We obtain the persistent current at full filling
in the form
\begin{equation}
I=-\frac{4\pi N}{\phi_0} \sum_{s=1}^{\infty} \ s \ \Gamma_{sN} \
\sin(2\pi\frac{\phi}{\phi_0}s) \ ,
\label{persistentfullresult}
\end{equation}
or equivalently in the form \eqref{persistentfullresult final}
which coincides with the result \eqref{currentVSfluxarb} for $a_j = 2 \Gamma_j$.

 In the rest of this section we discuss the physical meaning of the coefficients $\Gamma_N$, $\Gamma_{2N}$, etc., appearing in expression \eqref{persistentfullresult final}. Since $a_j = 2 \Gamma_j$, the meaning of the coefficients $a_N$, $a_{2N}$, etc., is the same and does not need an extra discussion.

According to the definition \eqref{hoppingcrystal},
in the infinite 1D crystal the matrix element $\Gamma_N$ represents the hopping amplitude of a single hop from the Wannier state at site $l$
to the Wannier state at site $l + N$. In the ring with $N$ lattice sites the site $l + N$ is a periodic replica of site $l$,
which means the same site. Appearance of $\Gamma_N$ in expression \eqref{persistentfullresult final} therefore implies,
that the electron in the ring with $N$ sites can make a single hop from the Wannier state at site $l$ to its periodic replica at site $l+N$.
In other words, the electron hops from site $l$ to the same site by making one round around the ring, and $\Gamma_N$ is the
hopping amplitude of the process. Note that we speak about hopping in the ring, but in terms of the Wannier states of the infinite crystal
(c.f. Eq. \ref{hoppingcrystal}).

Similarly, the coefficient $\Gamma_{2N}$ describes the process
in which the electron hops from site $l$ to the same site by making two rounds around the ring.
It follows from relation \eqref{Fouriercoefprop} and equation $a_j = 2 \Gamma_j$,
that $\Gamma_{2N}$ is much smaller than $\Gamma_{N}$. Its contribution to the right hand side of
Eq. \eqref{persistentfullresult final} is therefore negligible. The meaning of the coefficients $\Gamma_{3N}$, $\Gamma_{4N}$, etc., is analogous and their contribution is negligible even more.

We can summarize the result \eqref{persistentfullresult final} as follows. The persistent current in the insulating ring with $N$ lattice sites,
 derived in terms of the crystal Wannier functions, is due to the fact that
the electron at the ring site $l$ is allowed to make a single hop from site $l$ to its periodic replicas $l \pm N$.
(The sign minus is allowed as well due to the symmetry reason.)

The process in which an electron at the ring site $l$ hops from site $l$ to the same site by making one round around
the ring may seem meaningless, because
such process does not exist in the basis of the ring Wannier functions. In that basis the energy spectrum $\varepsilon_n(\phi)$ is given by
equation \eqref{eigenenergy2}. As shown in appendix B, equation \eqref{eigenenergy2} can be rewritten for odd $N$ as
\begin{equation}
\varepsilon_{n}(\phi) =
\sum_{j=-\frac{N-1}{2}}^{\frac{N-1}{2}}\ e^{-i{k_n(\phi)}ja}T_j \ ,
\label{eigenenergy2 2}
\end{equation}
and for even $N$ as
\begin{equation}
\varepsilon_{n}(\phi)
= \sum_{j=- \frac{N}{2}+1}^{\frac{N}{2}}\ e^{-i{k_n(\phi)}ja}T_j \ ,
\label{eigenenergy2 22}
\end{equation}
where $T_{-j} = T^*_j$. According to the definition \eqref{Tlprimel}, the matrix element $T_j$ represents the amplitude
of hopping from the ring Wannier state $w_l(x)$ to the ring Wannier state $w_{l^{\prime}}(x)$, with $j = l - l^{\prime}$
being the hopping length. The maximum hopping length between the ring Wannier states,
allowed by equations \eqref{eigenenergy2 2} and \eqref{eigenenergy2 22}, is $j = \pm {\rm Int} (N/2)$ only,
and the current in the insulating ring (Eq. \ref{persistentfull}) is due to the fact that the ring Wannier states depend on magnetic flux.
Hopping from $l$ to $l \pm N$, $l \pm 2N$, etc., emerges after the ring Wannier states are
expressed (see Eq. \ref{wanierringcrystal}) through the flux-independent crystal Wannier states.
This provides a different picture of the same physics.

\section{V. The LCAO approximation, application to the 1D lattice with rectangular potential wells}

\begin{figure}[t]
\vspace{-0.0cm} \centerline{\includegraphics[clip,width=0.95\columnwidth]{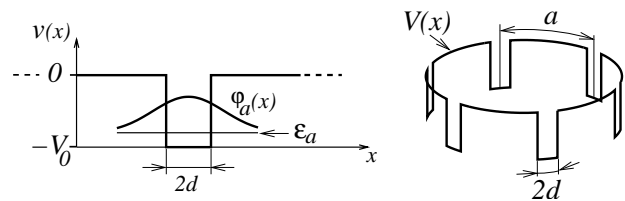}}
\vspace{-0.15cm} \caption{Left: One-dimensional model of a single isolated atom, described by
equation
\eqref{schr rov atomarna}. The atomic potential, $v(x)$, is modeled by the potential
well of width $2d$, embedded between two infinitely-thick potential barriers of height $V_0$.
Right: The potential $V(x)$ in a ring-shaped periodic lattice composed of $N$ one-dimensional
 atoms ($N=6$). The lattice period is $a$,
the ring circumference is $L = Na$. In the numerical calculations of Sect. V, the parameters of the ring are chosen to emulate the GaAs ring with conduction electrons
subjected to the periodic potential of $N$
quantum dots: the electron effective mass is $m = 0.067m_0$, the period of the quantum-dot lattice
is $a=40$nm, the ring length is $L=Na$, and the size and depth of the potential well (quantum dot)
are $2d=15$nm and $V_0=18$meV, respectively. For these parameters the potential well supports one bound state, the ground state with
 energy $\varepsilon_a = - 10.4$meV. The ground-state orbital $\varphi_a(x)$ is sketched schematically.} \label{Fig:1Dlatticemodel}
\end{figure}

To evaluate the persistent current (Eq. \ref{persistentfullresult final}) for a specific crystal lattice, we need to determine the crystal Wannier functions of the lattice. As an example we consider the ring-shaped 1D lattice in Fig. \ref{Fig:1Dlatticemodel}.
The lattice is composed of the artificial 1D
 atoms with a single atomic level $\epsilon_a$.
The atomic orbital $\varphi_a(x)$ and energy $\epsilon_a$ in the
isolated atom positioned at $x=0$ obey the Schr\"odinger equation
\begin{equation}
\label{schr rov atomarna} \left[-\frac
{\hbar^2}{2m}\frac{d^2}{dx^2} + v(x)\right]\varphi_a(x) \ = \
\epsilon_a \varphi_a(x) \ ,
\end{equation}
where $v(x)$ is the atomic potential modeled as a rectangular potential well centered at $x =0$ (figure \ref{Fig:1Dlatticemodel}, left).

 In accord with the general recipe of Sect. II, assume that the ring with $N$ atoms (figure \ref{Fig:1Dlatticemodel}, right) is periodically
 repeated as shown in figure \ref{Fig:N-atomicring}. This repetition gives rise to the infinite 1D crystal with potential $V(x)$
 which is
periodic with period $a$ and ranges from $x = - \infty$ to $x = \infty$.
Evidently, $V(x)$ is a sum of all isolated atomic potentials, that means
\begin{equation}
\label{infinite periodic potential} V(x)\ = \ \sum_{l=-\infty}^{\infty}\
v(x-la) \ ,
\end{equation}
where $la$ is the position of the $l$-th lattice point. The Bloch function $\Phi_{k}(x)$ and Bloch
energy $\epsilon(k)$ of the electron moving in such periodic potential can be expressed analytically in
 the approximation of localized atomic orbitals (LCAO). The Bloch function is approximated by the LCAO ansatz \cite{Ashcroft}
\begin{equation}
\Phi_{k}(x) \ = \
C \sum_{l=-\infty}^{\infty}\ e^{ikla}\varphi_a(x-la) \ ,
\label{Blochapprox}
\end{equation}
where $\varphi(x-la)$ is the atomic orbital at the lattice point $la$ and $C$ is the normalization constant (given below). The Bloch energy
$\epsilon(k)$ can be expressed as
\begin{equation}
\label{meanvalueof Blochenergy}
\epsilon(k) =  \langle\Phi_{k}\vert \hat H \vert
\Phi_{k}\rangle / \langle \Phi_{k}\vert\Phi_{k}\rangle,
\end{equation}
where
\begin{equation}
\label{hamiltoniannekonecnehokrystalu} \hat H = -\frac
{\hbar^2}{2m}\frac{d^2}{dx^2} + V(x), \ \ \ V(x)\ = \ \sum_{l=-\infty}^{\infty}\
v(x-la) \ .
\end{equation}
If we use the LCAO ansatz \eqref{Blochapprox}, a simple textbook
 calculation \cite{Ashcroft} gives the formula
\begin{equation} \label{Blochenergy-LAO-exact}
\epsilon(k)=\epsilon_a  - \frac {
\gamma_{0}+
2 \sum_{j=1}^{\infty} \gamma_{j}  \cos{(kja)}}{1 + 2
\sum_{j=1}^{\infty} \alpha_{j}  \cos{(kja)}} ,
 \
\end{equation}
where
\begin{equation} \label{wavefunctionoverlap}
\alpha_{j}=
\int_{-\infty}^{\infty} dx \varphi_a(x-ja)
 \varphi_a(x) ,
\end{equation}
\begin{equation} \label{interactionwavefunctionoverlap}
\gamma_{j}= -
\int_{-\infty}^{\infty} dx \varphi_a(x-ja) {V^{\prime}}(x)
 \varphi_a(x) , \quad \gamma_{0} = \gamma_{j=0},
\end{equation}
and
\begin{equation} \label{perturbation-interaction}
{V^{\prime}}(x) = V(x) - v(x) ,
\end{equation}
with $v(x)$ being the atomic potential at the lattice site $l = 0$. In Appendix C we express
$\alpha_{j}$ and $\gamma_{j}$ analytically for the model atomic lattice of Fig. \ref{Fig:1Dlatticemodel}.

Using the general recipe of section II,
we obtain the spectrum of the ring, $\epsilon_n(\phi)$, directly from the Bloch energy
\eqref{Blochenergy-LAO-exact}. It reads
\begin{equation}
\epsilon_n(\phi)=\epsilon_a -
\frac {\gamma_{0}+ 2 \sum_{j=1}^{\infty} \gamma_{j}  \cos{[k_{n}(\phi)ja]}}{1 + 2
\sum_{j=1}^{\infty} \alpha_{j}  \cos{[k_{n}(\phi)ja]}} ,
\label{Natomringenergy-LAO-exact}
\end{equation}
where $k_n(\phi) \equiv \frac{2 \pi}{Na}(n + \frac{\phi}{\phi_{0}})$.
One can substitute the spectrum \eqref{Natomringenergy-LAO-exact} into the persistent current formula
\eqref{eq:prud1} and one can evaluate the sum on the right hand side of Eq. \eqref{eq:prud1} numerically. We will refer to this approach as to the
numerical LCAO approach. The approach provides a straightforward numerical evaluation of persistent current in insulating rings
($N_e = N$) as well as in conducting rings ($N_e < N$). It does not provide understanding of the effect, however, it is useful for verification of
analytical results.

We
mainly want to verify our result for insulating rings, the formula \eqref{persistentfullresult final}. For a meaningful comparison with the numerical LCAO approach, the hopping amplitude $\Gamma_j$ in the formula \eqref{persistentfullresult final} has to be determined in the LCAO approximation.

We start by specifying the constant $C$ in the ansatz \eqref{Blochapprox}. Setting Eq. \eqref{Blochapprox}
into the normalization condition \eqref{blochCrystalNorm1} we find
\begin{equation}
\frac{1}{\mid C \mid^2}
=
1+2\sum_{j=1}^{\infty} \alpha_j \cos jka  \equiv N(k) \ .
\label{normconstC}
\end{equation}
The same result is obtained if we set Eq. \eqref{Blochapprox} into the orthogonality condition \eqref{blochCrystalNorm2}.

The crystal Wannier functions are defined by equation \eqref{wanierCrystal}. We set into equation \eqref{wanierCrystal}
the ansatz \eqref{Blochapprox} with $C = 1/\sqrt{N(k)}$.
We obtain the crystal Wannier function in the form \cite{Kohn2,Andreoni}
\begin{equation}
W_l(x)=\sum_{l'=-\infty}^{\infty}\ c_{l'-l}\ \varphi_a(x-l'a) \ ,
\label{wanier_by_atomic}
\end{equation}
where
\begin{equation}
c_{n}=\frac{a}{2\pi}\int_{-\frac{\pi}{a}}^{\frac{\pi}{a}}
\frac{\cos{(kna)}}{\sqrt{N(k)}}\ dk \ .
\label{wanier_coeficient}
\end{equation}
Expression \eqref{wanier_by_atomic} is the crystal Wannier function in the LCAO approximation. It is easy to verify that the expression  \eqref{wanier_by_atomic} fulfills the orthogonality condition  \eqref{blochCrystalNorm2} exactly.

If we express the crystal Wannier functions in Eq. \eqref{hoppingcrystal} by means of equation \eqref{wanier_by_atomic},
we find that
\begin{equation}
{\it \Gamma}_j=\sum_{n=-\infty}^{\infty}\sum_{n^{\prime}=-\infty}^{\infty}
c_{n-j}\ c_{n^{\prime}}\left(\varepsilon_a\alpha_{n-n^{\prime}}-\gamma_{n-n^{\prime}}\right) \ .
\
\label{G_by_gamma}
\end{equation}
Equation \eqref{G_by_gamma} is the hopping amplitude $\Gamma_j$ in the LCAO approximation.
An alternative form of Eq. \eqref{G_by_gamma} is the equation
\begin{eqnarray}
\Gamma_{j}
&=&
\frac {a}{2\pi}\int_{-\pi/a}^{\pi/a} dk \cos(jka)
\nonumber
\\
&\times&
\left[ \epsilon_a  - \frac {
\gamma_{0}+
2 \sum_{j'=1}^{\infty} \gamma_{j'}  \cos{(kj'a)}}{1 + 2
\sum_{j'=1}^{\infty} \alpha_{j'}  \cos{(kj'a)}} \right] \ ,
\nonumber
\\
\label{G_by_gamma 2}
\end{eqnarray}
obtained by setting the Bloch energy \eqref{Blochenergy-LAO-exact} into the Fourrier transformation \eqref{hoppingcrystal 2}.

We also note that the Bloch energy \eqref{Blochenergy-LAO-exact} is an alternative form of the expression \eqref{Blochenergy_wanier} with coefficients $\Gamma_j$
given by equation \eqref{G_by_gamma}, since
both forms follow from the same LCAO approximation. Proof of coincidence of both forms is straightforward, but lengthy
and therefore not shown.

The LCAO expression \eqref{G_by_gamma} together with expressions \eqref{wanier_coeficient} and \eqref{normconstC} are still
complicated for analytical calculations. Fortunately, the expressions can be simplified, if we take into account
that the coefficients $\alpha_j$ and $\gamma_j$ decay with increasing $j$ exponentially. Exponential decay is
due to the fact that the atomic orbitals $\varphi_a$ are exponentially localized at the lattice sites, which is the case for any realistic
orbital.
If the localization is strong enough,
then $1 \gg \mid \alpha_1 \mid \gg \mid \alpha_2 \mid \gg \dots$ and $1 \gg \mid \gamma_1 \mid \gg \mid \gamma_2 \mid \gg \dots$.
In that limit the expressions \eqref{wanier_coeficient} and \eqref{G_by_gamma} are calculated in appendix D. The results are
\begin{equation}
c_n
\simeq
 (-\alpha_1)^{n} \mid A_n \mid \ , \ \ \ n \geq 0 \ , \ \ \ c_{-n} = c_n \ ,
\label{c_n approximate}
\end{equation}
where $A_0 = 1$ and
\begin{equation}
A_n=
(-1)^n
\frac{1}{2^{2n}} \ \binom{2n}{n} \simeq  (-1)^n \frac{1}{\sqrt{\pi n}} \ , \ \ n \geq 1 \ ,
\label{expansion_coef2 maintext}
\end{equation}
and
\begin{equation}
{\it \Gamma}_j
\simeq -\gamma_1 (-\alpha_1)^{(j-1)} \ , \ \ \ \ j \geq 1.
\label{G_by_gamma_fin2A maintext}
\end{equation}
Expressing $\Gamma_j$ in equation \eqref{persistentfullresult final} by means of Eq. \eqref{G_by_gamma_fin2A maintext} and keeping
only the term with $j=N$ we obtain the result
\begin{equation}
I(\phi)
=
-\ \frac{2\pi}{\phi_0} \ \gamma_1 \ N \ (-\alpha_1)^{(N-1)}
\sin\left(2\pi\frac{\phi}{\phi_0}\right) \ .
\label{currentfull_aprox_wanier}
\end{equation}
The last equation is the persistent current in the insulating ring in the LCAO approximation.
The following features of the result \eqref{currentfull_aprox_wanier} are worth stressing.

For positive $\alpha_1$ equation \eqref{currentfull_aprox_wanier} predicts the exponentially decaying current with alternating sign, $I(N) \sim (-\alpha_1)^{(N-1)}$. Just this result holds for the ring in figure \ref{Fig:1Dlatticemodel}. Indeed,
$\alpha_1$ is positive for a positive $\varphi_a(x)$ such like the ground state orbital in figure \ref{Fig:1Dlatticemodel}, which follows from definition \eqref{wavefunctionoverlap} as well as from the final expression for $\alpha_j$ (appendix C).

For negative $\alpha_1$ the dependence $I(N) \sim (-\alpha_1)^{(N-1)}$ implies
the exponentially decaying current with fixed sign. For the square-well model in figure \ref{Fig:1Dlatticemodel}
the coefficient $\alpha_1$ is negative for the first excited bound state (not shown in the figure), because
the atomic orbital of the state is an antisymmetric function of $x$ with respect to the well center.
Thus, if the atomic level $\varepsilon_a$ and atomic orbital $\varphi_a(x)$ belong to the first excited bound state,
the sign of the current is the same for any $N$.

Equations \eqref{perzistentny vysledok_tight_binding odd}
and \eqref{perzistentny vysledok_tight_binding even} show that the persistent current in the conducting ring ($N_e < N$) exhibits the alternating sign
in dependence on the parity of $N_e$. This result is independent on the specific properties of the conduction band.
If the ring is insulating ($N_e = N$), the result \eqref{currentfull_aprox_wanier} shows that the sign of the current with respect to $N$ is either alternating or fixed, depending on the properties of the valence band. More precisely, this statement holds for the model ring in figure \ref{Fig:1Dlatticemodel}. We will see in the next section that the sign behavior can be even more complicated in rings made of the real band insulators.

Comparison of equation \eqref{currentfull_aprox_wanier} with Eqs. \eqref{perzistentny vysledok_tight_binding odd}
and \eqref{perzistentny vysledok_tight_binding even} also shows another important difference between the insulating and conducting rings.
In the former case the current decays with $N$ exponentially, in the latter case it decays like $1/N$.

\begin{figure}[t]
\centerline{\includegraphics[clip,width=0.9\columnwidth]{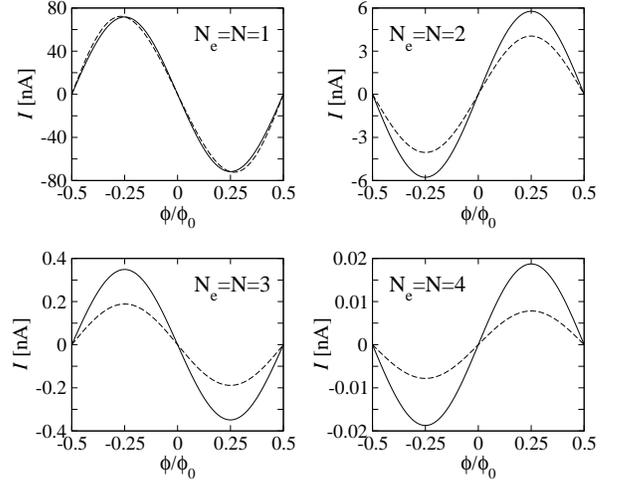}} \vspace{-0.15cm}
\caption{Persistent current in the insulating 1D ring as a function of magnetic flux for various ring lengths $N$.
The ring is the periodic lattice with $N$ single-level potential wells (Fig. \ref{Fig:1Dlatticemodel}) which contains
$N_e$ spinless electrons. It is insulating due to full filling ($N_e = N$).
The full lines are the results of the LCAO formula \eqref{currentfull_aprox_wanier} and the dashed lines are the results
of the numerical LCAO approach. The ring parameters used in these calculations are specified in the caption of figure \ref{Fig:1Dlatticemodel}. The coefficients $\alpha_j$ and $\gamma_j$ in Eqs. \eqref{currentfull_aprox_wanier} and \eqref{Natomringenergy-LAO-exact} are given by expressions presented in appendix C.
} \label{current in N-atomic ring versus flux}
\end{figure}

\begin{figure}[t]
\centerline{\includegraphics[clip,width=1.0\columnwidth]{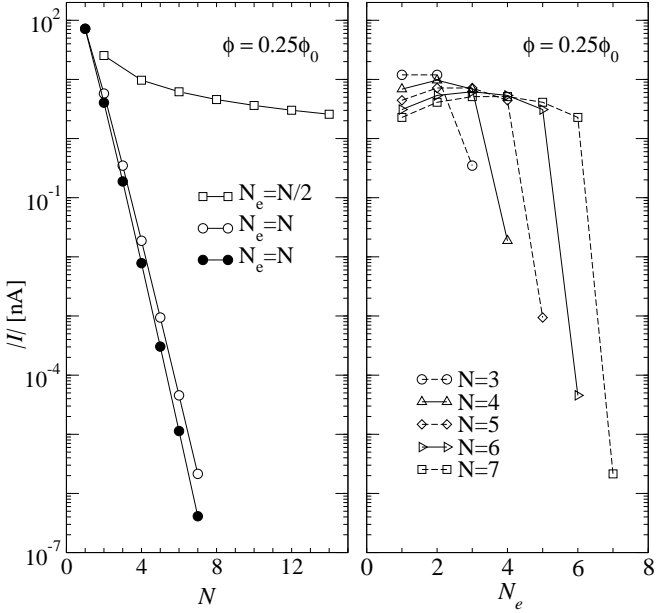}} \vspace{-0.15cm}
\caption{Persistent  current in the same ring as in the preceding figure for magnetic flux $\phi =
0.25\phi_0$. The left panel shows the dependence on $N$ for the insulating ring ($N_e=N$) and
conducting ring at half filling ($N_e=N/2$).
The right panel shows the dependence on $N_e$ for
various values of $N$. In both panels, the data for $N_e=N$ shown by open symbols were obtained by means of the LCAO expression
\eqref{currentfull_aprox_wanier} and the data for partial filling ($N_e<N$) were obtained by means of expressions \eqref{perzistentny vysledok_tight_binding odd}
and \eqref{perzistentny vysledok_tight_binding even}. The full circles in the left panel show the data for $N_e=N$,
obtained using the numerical LCAO approach. Note that only the size of the current, $\mid  I \mid$, is shown: for all presented data the sign of the current varies with varying parity of $N_e$.} \label{current
in N-atomic ring versus Ne}
\end{figure}

Origin of all these differences is easy to see. Equations \eqref{perzistentny vysledok_tight_binding odd}
and \eqref{perzistentny vysledok_tight_binding even} depend only on the hopping amplitude $\Gamma_1$, while the result
\eqref{currentfull_aprox_wanier} is determined by the hopping amplitude $\Gamma_N $. The amplitude
$\Gamma_N $ involves overlap of the Wannier functions $W_l(x)$ and $W_{l+N}(x)$. If we express the coefficient $c_{l'-l}$ in
equation \eqref{wanier_by_atomic} by means of equations \eqref{c_n approximate} and \eqref{expansion_coef2 maintext}, we
obtain
\begin{equation}
W_l(x) \simeq \sum_{l'=-\infty}^{\infty}\ \mid A_{\mid l'-l \mid } \mid (-\alpha_1)^{\mid l'-l \mid} \mid \ \varphi_a(x-l'a) \ .
\label{wanier_by_atomic approx}
\end{equation}
If we keep $x = na$ and $l=0$ for simplicity, we obtain
\begin{equation}
W_0(na) \simeq \frac{1}{\sqrt{\pi \mid n \mid}} (-\alpha_1)^{\mid n \mid}  \varphi_a(0) \ , \ \ \mid n \mid \geq 1 \ .
\label{wanier_by_atomic approx approx}
\end{equation}
It can be seen that the Wannier function decays with distance from site $l=0$ exponentially \cite{comment_Wannierfucntions} and
even shows the alternating sign for positive $\alpha_1$.
All these properties are reflected by persistent current at full filling.

Finally, we calculate the persistent current quantitatively for the numerical parameters
in figure \ref{Fig:1Dlatticemodel}. These parameters emulate the 1D GaAs ring modulated by a periodic lattice of $N$
quantum dots. In such lattice, the discrete energy level of the quantum dot splits into the energy band. Therefore,
the ring behaves at full filling ($N_e = N$) like a band insulator and at partial filling ($N_e < N$) as a metal.

The figure \ref{current in N-atomic ring versus flux} shows the persistent current in the insulating ring as a function of
 magnetic flux for various $N$. The results evaluated by means of
the LCAO formula \eqref{currentfull_aprox_wanier} are compared with the results obtained by the numerical LCAO approach.
The numerical LCAO approach nicely confirms basic features of the formula \eqref{currentfull_aprox_wanier} including the alternating sign of the current.
The quantitative difference between the numerical and analytical LCAO data could be suppressed
by choosing the ring with a more strongly localized atomic orbitals.

The dependence on the ring size is presented in figure \ref{current in N-atomic ring versus Ne}.
The left panel shows the persistent current as a
function of $N$ in the insulating ring and in the conducting ring at half filling. The conducting
ring shows for large $N$ the decay $I \propto 1/N$, in the insulating ring the current decays
with $N$ exponentially. For example, the insulating ring with $N = 5$ supports the persistent current as small as $\sim
0.001$nA. Such small persistent currents are detectable by sensitive
techniques \cite{Bleszynski,Bleszynski2,Bleszynski3}.

The right panel of figure
\ref{current in N-atomic ring versus Ne} shows the transition from the conducting state to the
insulating state at $N_e=N$, achieved by varying $N_e$ for fixed $N$. Experimentally, $N_e$ could
be varied say by using a suitably designed metallic gate, while the quantum-dot lattice might be
realized either by means of the periodic array of gates or by producing the 1D ring with a
periodically alternating cross-section size.

\section{VI. Estimates of persistent current in rings made of real band insulators}

In the previous section, we have discussed the ring made of the artificial band insulator,
namely the 1D GaAs ring with a few conduction electrons subjected to the periodic quantum-dot
potential. Now we want to estimate persistent current in rings made of the real
crystalline band insulators (with fully occupied valence band and empty conduction band).
The problem is that such rings are three-dimensional and the 1D theory from the preceding text
is not applicable directly.

\begin{figure}[t]
\centerline{\includegraphics[clip,width=0.82\columnwidth]{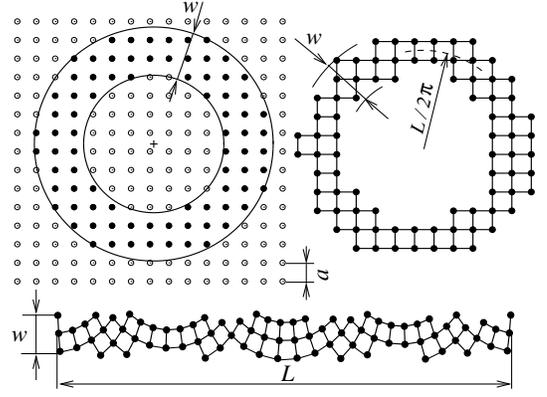}} \vspace{-0.15cm} \caption{The
top figure shows the 3D ring (on the right hand side) created from the cubic 3D atomic lattice (on
the left hand side) with the lattice period $a$. The ring of width $w$ is defined by two concentric
circles with the center positioned at random. The resulting ring is a 3D cluster of atoms which can
no longer be viewed as the 1D lattice with period $d$. The bottom figure shows the ring
artificially linearized by means of elastic deformation. Repeating the linearized ring with period
$L$ in the $x$ direction and with period $w$ in the $y$ direction, we obtain the infinite crystal.
In the limit $L \gg w$ the effect of the elastic deformation has to disappear and the Bloch states
in the obtained infinite crystal have to coincide with the electronic states in the real ring (top
right sketch). This allows us to use the equations \eqref{eq:prud1 3D} and \eqref{aproximacia
ohnuteho clusteru}. We call this model the model of the linearized 3D cluster.}
\label{Fig:inv-rubber}
\end{figure}

The problem is outlined in figure \ref{Fig:inv-rubber}.
The top right sketch in the figure is a 2D sketch of the 3D ring (or a hollow 3D cylinder) of width $w$,
created from the cubic 3D lattice of atoms. Along such
ring there is no periodicity with the lattice period $a$, there is only periodicity with $L$.
However, the theory in Sects. III - V was developed
for the 1D rings with periodicity $a$ along the ring circumference. The question is how to extend the 1D theory to be applicable to the
ring-shaped 3D cluster of atoms sketched in the figure \ref{Fig:inv-rubber}.

In principle, for $L \gg w$ it is reasonable to write the 3D version
of equation \eqref{eq:prud1} in the form
\begin{equation}
I(\phi)=-\frac{\partial}{\partial{\phi}}\sum_{n, k_z}\varepsilon_{n}(\phi,k_y=0,k_z),
\label{eq:prud1 3D}
\end{equation}
with the spectrum of the ring, $\varepsilon_{n}(\phi,k_y=0,k_z)$, determined as follows:
\begin{equation}
\varepsilon_{n}(\phi,k_y=0,k_z) = \varepsilon(k_x= \frac{2 \pi}{L}(n + \frac{\phi}{\phi_{0}}),k_y=0,k_z) ,
\label{aproximacia ohnuteho clusteru}
\end{equation}
where $\varepsilon(k_x,k_y,k_z)$ is the Bloch energy of the 3D crystal
created by periodic repetition of the linearized 3D cluster (Fig.\ref{Fig:inv-rubber}).
Note that the unit cell of the crystal is the linearized cluster.
The Bloch energy $\varepsilon(k_x,k_y,k_z)$ should therefore not be confused
with the Bloch energy of the crystal with unit cell $a^3$.
\emph{We call this approach the model of the linearized 3D cluster.}

The model also allows to rewrite for 3D rings all equations of Sect. III.
First of all, since the period $a$ along the ring no longer exists, it
has to be replaced by the period $L$ (replace $a$ by $L$ and $N$ by unity).
Second, the Bloch energy $\varepsilon(k_x)$ has to be replaced by the Bloch energy
$\varepsilon(k_x,k_y=0,k_z)$ just introduced. Third, the current \eqref{currentVSfluxarb} has to be summed
over the occupied states $k_z$. Clearly, the model of the linearized 3D cluster allows to avoid a full 3D treatment of the ring-shaped 3D cluster.
However, the model is still cumbersome for practical use because the Bloch spectrum
$\varepsilon(k_x,k_y,k_z)$ needs to be calculated for a nontrivial artificial crystal resulting from the periodic repetition of the linearized
cluster.

\begin{figure}[t]
\centerline{\includegraphics[clip,width=0.82\columnwidth]{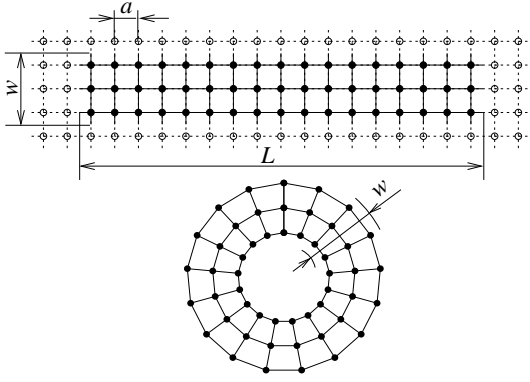}} \vspace{-0.15cm} \caption{The
3D ring of length $L$ and width $w$, created by elastic bending of the cubic 3D atomic lattice with
lattice period $a$. In the limit $w \ll L$ the effect of the elastic bending has to disappear and
the electronic states in such ring have to coincide with the Bloch states in the constituting 3D
crystal as described in the text (equation \ref{aproximacia ohnuteho krystalu 3D}). We call this
model the model of the elastically bent 3D crystal. Obviously, the resulting 3D ring shows for $w
\ll L$ an artificial periodicity with period $a$, which is not the case for the realistic 3D ring
(fig. \ref{Fig:inv-rubber}). } \label{Fig:rubber}
\end{figure}

A more simple approach is to consider the artificial 3D ring obtained by elastic bending of the bulk crystal,
as it is shown in figure \ref{Fig:rubber}. In this case we have instead of the formula
\eqref{eq:prud1 3D} the formula
\begin{equation}
I(\phi)=-\frac{\partial}{\partial{\phi}}\sum_{n,k_y,k_z}\varepsilon_{n}(\phi,k_y,k_z),
\label{eq:prud1 3D elast}
\end{equation}
where the energy spectrum $\varepsilon_{n}(\phi,k_y,k_z)$ is given by equation
\begin{equation}
\varepsilon_{n}(\phi,k_y,k_z) = \varepsilon(k_x= \frac{2 \pi}{Na}(n + \frac{\phi}{\phi_{0}}),k_y,k_z) ,
\label{aproximacia ohnuteho krystalu 3D}
\end{equation}
with $\varepsilon(k_x,k_y,k_z)$ being the Bloch energy in the bulk crystal before it is bent.
\emph{We call this approach the model of the elastically bent 3D crystal.} In this model there is an artificial periodicity
 with period $a$ along the ring.
In spite of this, the model can provide a reasonable quantitative information on how the persistent current in the 3D ring depends
 on the realistic band structure of the constituting bulk crystal. In the model,
 the persistent current at full filling is given by
equations \eqref{persistentfullresult final} and \eqref{hoppingcrystal 2} modified as follows. The Bloch energy
$\varepsilon(k_x)$ in the Fourrier transformation \eqref{hoppingcrystal 2} has to be replaced by the
Bloch energy $\varepsilon(k_x,k_y,k_z)$
and the current \eqref{persistentfullresult final} should be summed over the occupied states $k_y$ and $k_z$.
Practical calculations are relatively straightforward, because the Bloch spectrum of the bulk crystal,
$\varepsilon(k_x,k_y,k_z)$, is known for many insulators.

In what follows we apply the model of the elastically bent 3D crystal to the rings with thickness equal to the unit cell
of the bulk crystal. The energy dispersion of the ring, $\varepsilon_{n}(\phi,k_y=0,k_z=0)$, is extracted from the bulk crystal
dispersion $\varepsilon(k_x,k_y=0,k_z=0)$, calculated by microscopic methods \cite{Chadi-Cohen,Loehr}. The obtained persistent current
should be viewed as a minimum value estimate, because the thickness of realistic rings is much larger than the unit cell.

\begin{figure}[t]
\centerline{\includegraphics[clip,width=0.9\columnwidth]{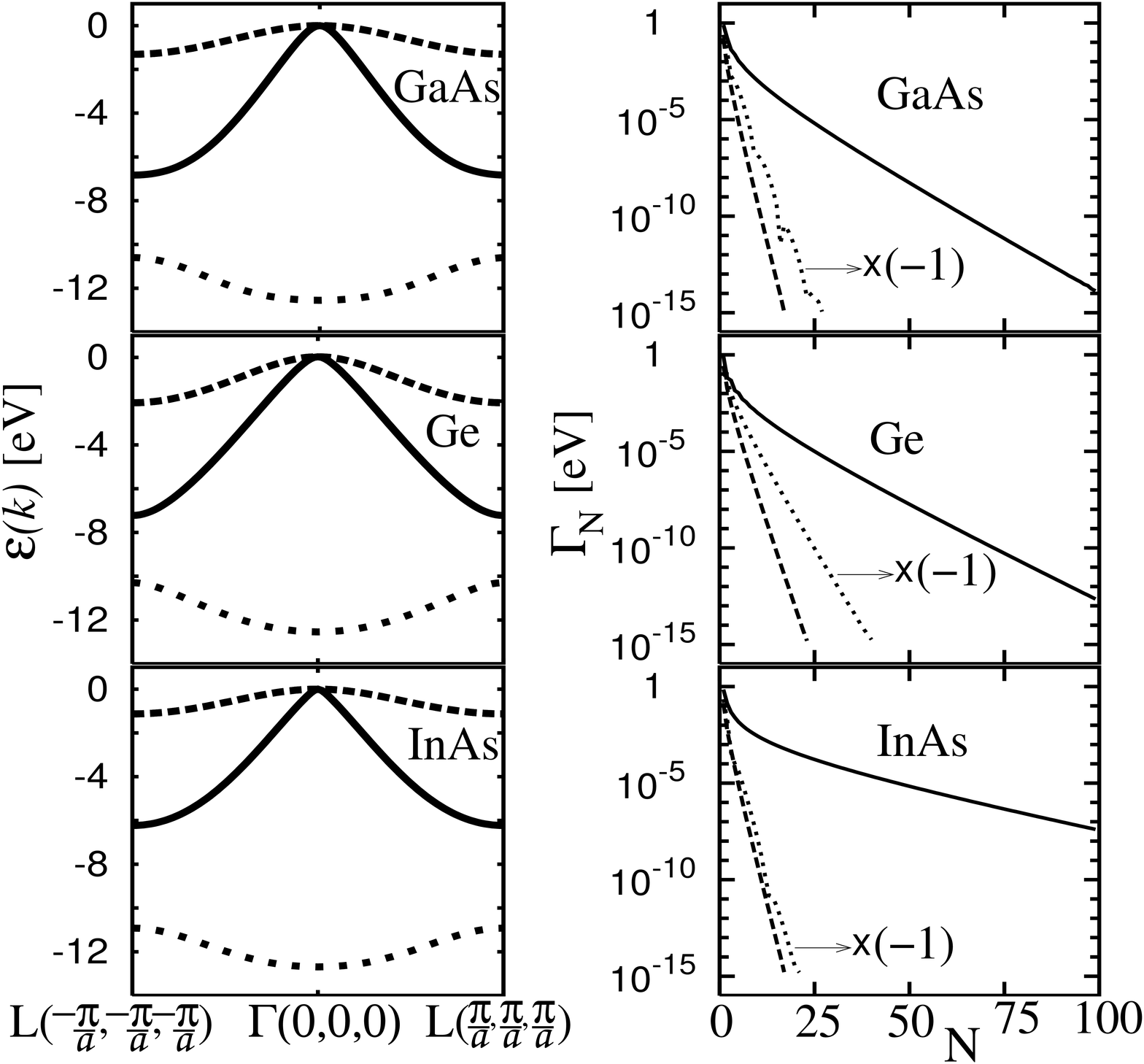}} \vspace{-0.15cm} \caption{The
panels in the right column show the valence bands of various band insulators like the GaAs, Ge, and
InAs. Specifically, the $\epsilon(k)$ dispersion curves of the light-hole band (full line),
double-degenerate heavy-hole band (dashed line), and lowest band (dotted line) are plotted for the
$k$-values on the line ($L[-\pi/a,-\pi/a,-\pi/a], \Gamma[0,0,0], L[\pi/a,\pi/a,\pi/a]$). The panels
in the left column show the hopping amplitude $\Gamma_N$, calculated separately for each
band: the $\epsilon(k)$ dependence of the band is Fourier-transformed by using the equation
\eqref{hoppingcrystal 2}. The full, dashed, a dotted lines show $\Gamma_N$ for the light-hole,
heavy-hole, and lowest bands, respectively. When compared with $\Gamma_N$ of the light-hole and
heavy-hole band , $\Gamma_N$ of the lowest
 band has opposite sign because of the opposite curvature of the $\epsilon(k)$ curve and the
 presented data should be multiplied by $-1$ as shown in the figure (but see the comment \cite{comment1}).
 } \label{bandsGaAsGeInAs}
\end{figure}

Using  standard methods \cite{Chadi-Cohen,Loehr}, in the figure \ref{bandsGaAsGeInAs} we numerically
reproduce the well-known band structure of the band insulators like the GaAs, Ge, and InAs.
Specifically, the figure summarizes the $\epsilon_k$ curves of the light-hole band, heavy-hole band
and lowest band, calculated for the line ($L[-\pi/a,-\pi/a,-\pi/a], \Gamma[0,0,0],
L[\pi/a,\pi/a,\pi/a]$) in the first Brillouin zone. We set each of these $\epsilon(k)$ curves
numerically into the Fourier transformation \eqref{hoppingcrystal 2} and we calculate for each band
the hopping amplitude $\Gamma_N$. The resulting $\Gamma_N$ dependencies are shown in the
figure \ref{bandsGaAsGeInAs}.

By means  of the figure \ref{bandsGaAsGeInAs} we estimate the persistent current as follows. We
form the ring by elastic bending of the crystal. We choose the ring orientation for which the $k$
states on the line ($L[-\pi/a,-\pi/a,-\pi/a], \Gamma[0,0,0], L[\pi/a,\pi/a,\pi/a]$) are directed
along the ring circumference. If the ring is one-dimensional (with thickness equal to a single unit
cell of the crystal), the persistent current is given by the 1D formula \eqref{persistentfullresult final}
with the hopping amplitude $\Gamma_N$ given by numerical values in the figure
\ref{bandsGaAsGeInAs}. As expected, $\Gamma_N$ decays with increasing $N$ exponentially, however, for
the light-hole band the decay is much slower than for other two bands. Therefore it is
sufficient to calculate the persistent current in the light-hole band. The results are shown in the
figure \ref{bandsGaAsGeInAs}. We recall that these results represent the persistent current carried
by electrons in the fully occupied valence band. The following features are worth to
point out.

\begin{figure}[t]
\centerline{\includegraphics[clip,width=0.95\columnwidth]{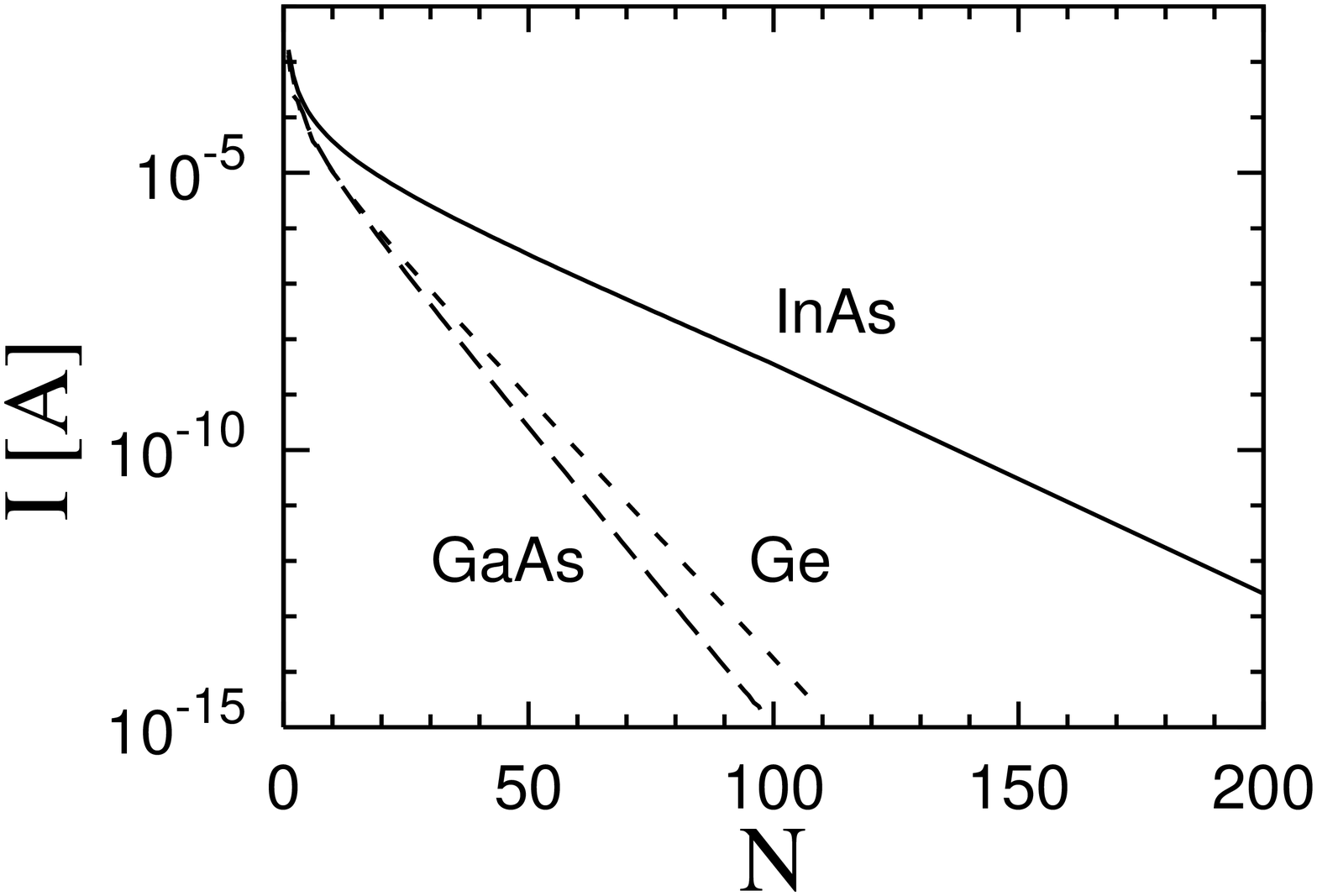}} \vspace{-0.15cm}
\caption{Persistent current versus the ring length $N$ in the ring made of the real crystalline
band insulator (GaAs, Ge, InAs). The persistent current is estimated by means of the 1D formula
\eqref{persistentfullresult final}, where $\Gamma_N$ is the hopping amplitude of the electron in the
light-hole band, calculated in the preceding figure. The lattice constants of the GaAs, Ge, and
InAs crystals are $\sim 0.6$nm, i.e., the maximum ring lengths considered in the figure are $L=Na
\sim 120$nm. The rings of such length could be realizable by means of advanced nanotechnology. }
\label{estimated persistent currents in GaAs Ge InAs}
\end{figure}

The persistent current in the insulating ring decreases with the increasing ring length
exponentially, but a careful choice of the insulating material allows to achieve the
persistent current of measurable value for a technologically realizable ring size. Indeed, the
maximum ring lengths considered in the figure \ref{estimated persistent currents in GaAs Ge InAs}
are close to the value $\sim 120$nm, in principle achievable by modern nanotechnology.
Concerning the amplitude of the persistent current, the value $\sim 0.0003$nA estimated for the
InAs ring of length $\sim 120$nm is very small, but in principle detectable say by the experimental
technique of Refs. \cite{Bleszynski,Bleszynski2,Bleszynski3}. The considered insulators are ordered
as GaAs, Ge and InAs for the persistent currents ordered increasingly. This suggests a
simple rule: the smaller the light-hole effective mass the larger the persistent current.

We conclude with a few remarks showing that the above 1D estimates hold roughly also for the 3D rings.
Similar calculations as in the figure \ref{bandsGaAsGeInAs} were performed for the $k$ values on
 the line  ($X[-2\pi/a,0,0], \Gamma[0,0,0], X[2\pi/a,0,0]$). We have found the energies $\epsilon(k)$
 and amplitudes $\Gamma_N$ (not shown) very similar to those in the figure \ref{bandsGaAsGeInAs},
 with the resulting persistent currents
being of the same order of magnitude as those in the figure \ref{estimated persistent currents in
GaAs Ge InAs}. Moreover, we have observed similar results when testing the $k$ values on the lines
($K, \Gamma, K$) and ($W, \Gamma, W$). The model of the elastically bent crystal thus provides
roughly the same persistent current independently on the fact which crystalline axis is chosen to
be directed along the ring circumference. This suggests that anisotropy of the valence band with
respect to the $\vec{k}$ direction does affect the resulting persistent
current dramatically. Therefore, the estimated values of the current should hold reasonably also for the realistic
ring-shaped crystals similar to the ring sketched in figure \ref{Fig:inv-rubber}.

\section{VII. Summary and concluding remarks}

We have analyzed theoretically persistent currents in rings made of band insulators. We started by formulating a recipe which
determines the Bloch states of a one-dimensional (1D) ring from the Bloch states of an
infinite 1D crystal created by the periodic repetition of the ring. Using the recipe, we have derived an expression
for the persistent current in a 1D ring made of an insulator with an arbitrary valence band $\epsilon(k)$.

To find an exact result for a specific insulator, we have considered a 1D ring represented by a periodic lattice of $N$ identical sites with a single value of the on-site energy. If the Bloch states in the ring are expanded
over a complete set of $N$ on-site Wannier functions, the discrete on-site energy splits into the
energy band. At full filling, the band emulates the valence band of the band insulator and the ring is insulating.

We have found that the ring carries
a persistent current equal to the product of $N$ and the derivative of the on-site energy with respect to the magnetic flux.
This current is not zero because the on-site Wannier function and consequently the on-site energy of each ring site depend on magnetic flux.

Further, we have expanded all $N$ Wannier functions of the ring over the basis
of Wannier functions of the constituting infinite 1D crystal. In terms of the crystal Wannier functions, the current at full
filling arises because the electron in the ring with $N$ lattice sites is allowed
to make a single hop from site $i$ to its periodic replicas $i \pm N$.
Alternatively, in terms of the ring Wannier functions the same current is due to the flux dependence of the Wannier functions basis, and
the longest allowed hop is from $i$ to $i \pm {\rm Int} (N/2)$ only.

We have eventually derived the crystal Wannier functions in the LCAO approximation,
and have expressed the persistent current at full filling by means of an exact formula.
The validity of the formula was verified by comparison with the numerical LCAO approach.
We have presented quantitative results for a 1D ring made of an artificial band insulator,
namely for a GaAs ring subjected to a periodic potential emulating the insulating lattice.

Moreover, we
have provided estimates for the rings made of real band insulators
(GaAs, Ge, InAs) with a fully filled valence band and empty conduction band. The current at full filling decays with the ring length exponentially due to the exponential decay of the Wannier function tails. In spite of that, it can be of measurable size.

We have ignored the effect of nonzero temperature. We expect
the persistent currents in rings made of the band insulators to be temperature-independent as long the
thermal energy $k_BT$ is much smaller than the energy gap separating the valence band from the conduction
band. For example, the energy gap in the InAs crystal is $\sim 400$meV. Hence, the zero-temperature
values of the persistent current, predicted for the insulating InAs ring, could persist even at
room temperature.

We have also ignored the electron-electron interaction. The
many-body studies  \cite{Mosko-PRB,Nemeth-PRB} suggest that in the 1D ring considered in
this paper the repulsive electron-electron interaction would only renormalize the electron
transmission through the potential barrier separating two potential wells. That means that the
electron-electron interaction would not essentially change our results.

Finally, we relate our results to the paper by Kohn \cite{Kohn1},
which defines the insulating state as a property of the many-body ground state (see also the review by Resta \cite{Resta}). Kohn's paper contains a discussion implicitly admitting
the existence of a negligibly small persistent current in the insulating ring [see the discussion of equations (2.23) - (2.31) in that work]. Kohn
points out the exponential localization of the electron states in the ring. However, he clearly states by writing his equation (2.30),
that the current is zero only approximately. Just this negligibly small current is calculated in our paper. Surprisingly, it can be of measurable size.

\section{Acknowledgements}
This work was supported by grants APVV-51-003505 and
VVCE-0058-07 from the Grant Agency APVV  and by grants VEGA-2/0633/09 and VEGA-2/0206/11
from the Grant Agency VEGA.
We thank Pavol Quittner for
important help with the mathematical derivation in Appendix D.

%

%

\section{Appendix A: Derivation of equation (\ref{e_k_n_finite}) from equation (\ref{e_k_n text})}

We start by writing Eq. \eqref{e_k_n text} as
\begin{equation}
\varepsilon_{n}(\phi) = \sum_{j'=0}^\infty a_{j'}
\cos \left(k_n(\phi)\ j'a\right)\  ,
\label{e_k_n append}
\end{equation}
Using substitution $j'=Ns+j$, where
$s=0,1,2, \dots$ and $j=0,1, \dots (N-1)$, we obtain
\begin{equation}
\varepsilon_{n}(\phi)=
\sum_{j=0}^{N-1}\sum_{s=0}^{\infty}
a_{sN+j}
\cos\left[k_n(\phi) (sN+j) a\right] .
\label{ring-ham-partial}
\end{equation}
Assuming that $N$ is odd we rewrite Eq. \eqref{ring-ham-partial} as
\begin{eqnarray}
\varepsilon_{n}(\phi)
&=&
\sum_{s=0}^{\infty} a_{sN} \cos\left[k_n(\phi) sN a\right]
\nonumber \\
&+&\sum_{j=1}^{M}
\sum_{s=0}^{\infty}
a_{sN+j}
\cos\left[k_n(\phi) (sN+j) a\right]
\nonumber \\
&+&
\sum_{j=M+1}^{N-1}
\sum_{s=0}^{\infty}
a_{sN+j} \cos\left[k_n(\phi) (sN+j) a\right] \ ,
\nonumber \\
\label{e_k_n1}
\end{eqnarray}
where $M = {\rm Int} (N/2)$. Equation \eqref{e_k_n1} can be rewritten as
\begin{eqnarray}
\varepsilon_{n}(\phi)
&=&
\sum_{s=0}^{\infty} a_{sN} \cos\left[k_n(\phi) sN a\right]
\nonumber \\
&+&\sum_{j=1}^{M}
\sum_{s=0}^{\infty}
\Bigl\{
a_{sN+j} \cos\left[k_n(\phi) (sN+j) a\right]
\Bigr .
\nonumber
\\
&+&
\Bigl .
a_{(sN+N-j)} \cos\left[ k_n(\phi) \left(sN+N-j \right) a \right ]
\Bigr\}
\ .
\nonumber
\label{e_k_n2}
\end{eqnarray}
Utilizing definition \eqref{k_n text} we find from Eq. \eqref{e_k_n2} the result
\begin{equation}
\varepsilon_{n}(\phi)=
\Omega_0 +
\sum_{j=1}^M
\Bigl\{
\Omega^{R}_j
\cos \left(k_n(\phi)\ ja\right)
+
\Omega^{I}_j
\sin \left(k_n(\phi)\ ja\right)
\Bigr \} \ ,
\label{e_k_n_result}
\end{equation}
where
\begin{eqnarray}
&&
\Omega_0(\phi)
=\sum_{s=0}^{\infty} a_{sN}\ \cos(2\pi\frac{\phi}{\phi_0}s) \ ,
\nonumber \\
&&
\Omega^{R}_j(\phi)=
a_j +
\sum_{s=1}^{\infty}\
\cos(2\pi\frac{\phi}{\phi_0}s)
\left[ a_{sN+j} + a_{sN-j} \right] \ ,
\nonumber \\
&&
\Omega^{I}_j(\phi)=
\sum_{s=1}^{\infty}\
\sin(2\pi\frac{\phi}{\phi_0}s)
\left[ a_{sN-j} - a_{sN+j} \right] \ .
\nonumber \\
\label{Omega odd}
\end{eqnarray}
We have sofar assumed odd $N$. For even $N$, derivation is similar and we find again the results \eqref{e_k_n_result}
and \eqref{Omega odd}, except that for $j = N/2$ we have instead of Eqs. \eqref{Omega odd} the equations
\begin{eqnarray}
&&
\Omega^{R}_{N/2}(\phi)
=\sum_{s=0}^{\infty}\
a_{sN+N/2}\ \cos(2\pi\frac{\phi}{\phi_0}s) \ ,
\nonumber \\
&&
\Omega^{I}_{N/2}(\phi)=-\sum_{s=1}^{\infty}\
a_{sN+N/2}\sin(2\pi\frac{\phi}{\phi_0}s) \ .
\nonumber \\
\label{Omega even}
\end{eqnarray}
Equation \eqref{e_k_n_result} coincides with Eq. (\ref{e_k_n_finite}).

\section{Appendix B: Various forms of equation (\ref{eigenenergy2})}

We first assume odd $N$. We can rewrite Eq. \eqref{eigenenergy2} as
\begin{equation}
\varepsilon_n(\phi)
=
T_0+ \sum_{j=1}^{M}\ e^{-i{k_n(\phi)}ja}T_j+\sum_{j=M+1}^{N-1}e^{-i{k_n(\phi)}ja}T_j,
\label{eigenenergy41}
\end{equation}
where $M = \rm Int (N/2)$. Further,
\begin{eqnarray}
\varepsilon_n(\phi)&=&
T_0+ \sum_{j=1}^{M}\ e^{-i{k_n(\phi)}ja}T_j+\sum_{j=1}^{M}e^{-i{k_n(\phi)}(N-j)a}T_{N-j}
\nonumber \\
&=&
T_0+\sum_{j=1}^{M}\ e^{-i{k_n(\phi)}ja}T_j+e^{i{k_n(\phi)}ja}T^*_j \ ,
\nonumber \\
\label{eigenenergy4}
\end{eqnarray}
\noindent
where we have used the definition \eqref{k_n text} and relations
\begin{equation}
T^*_j=T_{-j} \ ,
\quad \quad
T_{N-j}=e^{i2\pi\frac{\phi}{\phi_0}}T^*_j \ ,
\label{Tj_relations}
\end{equation}
which follow from Eq. \eqref{Tj} for real $\varepsilon_n(\phi)$.
For even $N$ we obtain in a similar way the equation
\begin{eqnarray}
\varepsilon_n(\phi)&=&
T_0+\sum_{j=1}^{M -1}\ e^{-i{k_n(\phi)}ja}T_j+e^{i{k_n(\phi)}ja}T^*_j
\nonumber \\
&+&
e^{-i{k_n(\phi)}\frac N2 a}T_{\frac N2}
\ .
\nonumber \\
\label{eigenenergy42}
\end{eqnarray}
\noindent
Equations \eqref{eigenenergy4} and \eqref{eigenenergy42} can be written in the unified form
\begin{equation}
\varepsilon_n(\phi) = \Lambda_0 + \sum_{j=1}^M
\Bigl\{
\Lambda^{R}_j \cos \left(k_n(\phi)\ ja\right)
+
\Lambda^{I}_j \sin \left(k_n(\phi)\ ja\right)
\Bigr \}
\ ,
\label{e_k_n_result_ring}
\end{equation}
where
\begin{eqnarray}
&&
\Lambda_0 = T_0 \ ,
\nonumber \\
&&
\Lambda^{R}_j=(T_j+T^{\ast}_j) \ ,
\nonumber \\
&&
\Lambda^{I}_j=\frac 1i(T_j-T^{\ast}_j) \
\nonumber \\
\label{Lambda odd}
\end{eqnarray}
for odd $N$ as well as even $N$, except that for even $N$ and $j = N/2$ we have instead of equations \eqref{Lambda odd} the equations
\begin{eqnarray}
&&
\Lambda^{R}_{N/2}=\frac 12 (T_j+T^{\ast}_j) \ ,
\nonumber \\
&&
\Lambda^{I}_{N/2}=\frac {1}{2i} (T_j-T^{\ast}_j) \ .
\nonumber \\
\label{Lambda even}
\end{eqnarray}
Equation \eqref{e_k_n_result_ring} is formally equivalent with Eq. \eqref{e_k_n_result}. Coincidence of both equations becomes obvious if we realize  that equations \eqref{Lambda odd} and \eqref{Lambda even} coincide with Eqs. \eqref{Omega odd} and \eqref{Omega even}. To observe this coincidence, it is sufficient
to express $T_j$ in equations \eqref{Lambda odd} and \eqref{Lambda even}
by means of expression  \eqref{Tgeneral}, and to use the equations $a_0=\Gamma_0$
and $a_j=2\Gamma_j$.

\section{Appendix C: Analytical formulas for $\alpha_j$ and $\gamma_j$}

For the square-well potential in figure \ref{Fig:N-atomicring} the atomic orbitals $\varphi_a(x-la)$
can be found easy and the integrals
 \eqref{wavefunctionoverlap}
and \eqref{interactionwavefunctionoverlap} can be calculated analytically. For $j = 1, 2, \dots$ we obtain
\begin{equation}
\label{alphaNanalyt} \alpha_j\simeq\frac{\cos^2(A d)}{1+B d}e^{-B(ja-2d)}
\left[\frac{-4\epsilon_a}{V_0}+B(ja-2d)\right] \ ,
\end{equation}
\begin{equation}
\label{gammaNanalyt} \gamma_j\simeq V_0\frac{\cos^2(A d)}{1+B d}e^{-B(ja-2d)}
\left[\frac{-2\epsilon_a}{V_0}+2dB(j-1)\right] \ ,
\end{equation}
where
\begin{equation}
\label{AB} A=\sqrt{\frac{2m}{\hbar^2}\left(V_0+\epsilon_a\right)}, \quad \quad
B=\sqrt{\frac{2m}{\hbar^2}\left|\epsilon_a\right|} .
\end{equation}
In addition, for $j = 0$
\begin{equation}
\label{gamma0analyt} \gamma_0=V_0\frac{\cos^2(A d)}{1+B d}e^{-2B(a-2d)}
\frac{1-e^{-4Bd}}{1-e^{-2Ba}} \ .
\end{equation}
Expressions \eqref{alphaNanalyt} and \eqref{gammaNanalyt} are not exact (we have skipped some
exponentially small terms), but they are in excellent agreement with numerical integration of the
integrals \eqref{wavefunctionoverlap} and \eqref{interactionwavefunctionoverlap}. As expected,
the coefficients $\alpha_j$ and $\gamma_j$ decay with increasing $j$ exponentially. The simplest forms are
\begin{equation}
\gamma_j \simeq \gamma_1 \ e^{-(j-1)B a} \ , \ \ \
\alpha_j \simeq \alpha_1 \ e^{-(j-1)B a} \ , \ \ \ j\geq1 \ .
\label{gamma_alpha_specif 1}
\end{equation}

\section{Appendix D: Approximate formulas for coefficients $c_n$ and $\Gamma_j$}

\noindent
Equation \eqref{wanier_coeficient} defines $c_n$ by means of the integral which can be calculated analytically in a
certain limit. In what follows the calculation is presented for positive $n$ because $c_n=c_{-n}$.
We first express the normalization constant $\frac {1}{\sqrt{N(k)}}$ as
\begin{eqnarray}
\frac {1}{\sqrt{N(k)}}
&=&
\left[1+2\sum_{j=1}^{\infty} \alpha_j \cos jka \right]^{-\frac 12}
\nonumber \\
&=&
\sum_{s=0}^{\infty}\ A_s \left[2\sum_{j=1}^{\infty} \alpha_j \cos jka\right]^{s}\ ,
\label{norm_expansion 0}
\end{eqnarray}
where
\begin{equation}
 A_s=
(-1)^s
\frac{1}{2^{2s}} \ \binom{2s}{s} \ .
\label{expansion_coef}
\end{equation}
A closer inspection of the right hand side of Eq. \eqref{norm_expansion 0} shows, that the equation \eqref{norm_expansion 0} can be approximated as
\begin{equation}
\frac {1}{\sqrt{N(k)}}
\simeq
\sum_{s=0}^{\infty}\ A_s \left[2 \alpha_1 \cos ka\right]^{s}\ ,
\label{norm_expansion}
\end{equation}
if $\alpha_j$ falls with $j$ exponentially as in equation \eqref{gamma_alpha_specif 1}, and if $\alpha_1^2 \gg \alpha_2$.
We note that the latter condition takes the form $\alpha_1 \gg \exp(-Ba)$ for $\alpha_j$ derived in the preceding appendix.
In general, both conditions are fulfilled simultaneously if the localization of the atomic orbitals is strong enough in comparison with the
lattice constant $a$.

We set the expansion \eqref{norm_expansion} into the formula \eqref{wanier_coeficient} and we perform the integration. We get
\begin{equation}
c_{n}
\simeq
\sum_{s=0}^{\infty}\ A_s \left[2\alpha_1\right]^s\ I_{n,s}
\label{wanier_coeficient_app1} \ ,
\end{equation}
where
\begin{equation}
I_{n,s}=\frac{1}{\pi}\int_{-{\pi}}^{{\pi}}
\cos(nx)\cos^s(x) dx \ .
\label{RG_integral}
\end{equation}
It can be seen that
\begin{equation}
I_{n,s}=0 \ ,
\label{RG_integral_resB}
\end{equation}
if $s < n$ or if $(s-n)$ is odd. It can also be seen that

\begin{equation}
I_{n,s}=\frac {2}{2^{n+2l}}
\ \binom{n+2l}{l}
\ ,
\label{RG_integral_resA}
\end{equation}
if $s=n+2l$, where $l=0,1,\dots$ .

We set equations \eqref{RG_integral_resA} and \eqref{RG_integral_resB} into the equation
\eqref{wanier_coeficient_app1}. We get an approximate expression for $c_n$,
\begin{equation}
c_{n}
=
\left(-\alpha_1\right)^{n}
\sum_{l=0}^{\infty}\ A_{n+2l} \
\alpha_1^{2l} \
\binom{n+2l}{l}
\ .
\label{wanier_coeficient_app2}
\end{equation}
The last equation can be further rewritten as
\begin{eqnarray}
c_{n}
&=&
\left(-\alpha_1\right)^{n}\left | A_n\right |
\sum_{l=0}^{\infty}\  \frac {A_{n+2l}}{A_n} \
\alpha_1^{2l} \
\binom{n+2l}{l}
\nonumber
\\
&<&
\left(-\alpha_1\right)^{n}\left | A_n\right |
\sum_{l=0}^{\infty}\alpha_1^{2l} \
\binom{n+2l}{l}
  \ ,
\label{wanier_coeficient_app3}
\end{eqnarray}
where the right hand side is a properly chosen upper limit (because $A_{n+2l}/A_n<1$).

If we use the relation
\begin{equation}
\sum_{l=0}^{\infty}
 (\alpha_1^2)^l\binom{n+2l}{l}
=
\frac 1{\sqrt{1-4\alpha_1^2}}\left(\frac 2{1+\sqrt{1-4\alpha_1^2}} \right)^n \ ,
\label{formula_sum}
\end{equation}
the inequality \eqref{wanier_coeficient_app3} reads
\begin{equation}
c_{n}
<
\left(-\alpha_1\right)^n
\left| A_n \right|
\frac 1{\sqrt{1-4\alpha_1^2}}
\left(\frac 2{1+\sqrt{1-4\alpha_1^2}} \right)^n
\ .
\label{wanier_coeficient_up}
\end{equation}
For $\alpha_1\ll 1/2$ it can be further approximated as
\begin{equation}
c_{n}
<
\left(-\alpha_1\right)^n
\left| A_n \right|
\left(1+\alpha_1^2 \right)^n
\label{wanier_coeficient_up_small}
\end{equation}
We return again to the equation \eqref{wanier_coeficient_app2} and we replace its right hand side by a properly chosen lower limit:
\begin{eqnarray}
c_n
&>&
\left(-\alpha_1\right)^n
\left| A_n \right|
\sum_{l=0}^{n}\binom{n}{l}\left(\alpha_1\right)^{2l}
\nonumber
\\
&>&
\left(-\alpha_1\right)^n
\left| A_n \right|
\left(1+\alpha_1^2\right)^n
\ ,
\nonumber
\\
\label{wanier_coeficient_down}
\end{eqnarray}
where we have utilized the inequality
\begin{equation}
\frac{A_{n+2l}}{A_n}\binom{n+2l}{l}
>
\binom{n}{l} \ .
\end{equation}
Notice now that the right hand sides of the inequalities \eqref{wanier_coeficient_up_small} and
 \eqref{wanier_coeficient_down} coincide. This implies that
\begin{equation}
c_{n} =
\left(-\alpha_1\right)^n
\left(1+\alpha_1^2\right)^n
\left| A_n \right| \simeq \left(-\alpha_1\right)^n \left| A_n \right|
\ .
\label{wanier_coeficient_final_a}
\end{equation}
%
\noindent

Now we attempt to find an approximate expression for the coefficient ${\it\Gamma}_j$.
Equation \eqref{G_by_gamma} can be rewritten by means of substitution $n^{\prime}-n=m$ as
\begin{equation}
{\it \Gamma}_j=
\sum_{m=-\infty}^{\infty}\ (\varepsilon_a\alpha_m-\gamma_m) \sum_{n=-\infty}^{\infty}c_nc_{j+m+n}
\ .
\label{G_by_gamma_rearanged}
\end{equation}
Since $c_n=c_{-n}$ , $\alpha_n=\alpha_{-n}$, and
$\gamma_n=\gamma_{-n}$, we obtain
\begin{eqnarray}
&
{\it \Gamma}_j&=
\left(\varepsilon_a\alpha_0-\gamma_0\right)
\Bigl [c_0 c_j + \sum_{n=1}^{\infty}c_n
\left(c_{j+n}+c_{j-n}\right)\Bigr ]
\nonumber
\\
&+&
\sum_{m=1}^{\infty}
\left(\varepsilon_a\alpha_m-\gamma_m\right)
\Bigl [
c_0 \left(c_{j+n}+c_{j-n}\right) \Bigr .
\nonumber
\\
&+&
\Bigl.
\sum_{n=1}^{\infty}c_n\left(c_{j+m+n}+c_{j+m-n}
+c_{j-m+n}+c_{j-m-n}\right)
\Bigr ]
\ .
\nonumber
\\
\label{G_by_gamma_rearanged1}
\end{eqnarray}
\noindent
Owing to the exponential dependence \eqref{wanier_coeficient_final_a}, the last equation can be simplified as
\begin{equation}
\Gamma_j \simeq \sum_{m=0}^{j}\left(\varepsilon_a\alpha_m-\gamma_m\right)
\sum_{n=0}^{j-m}c_nc_{j-m-n}
\ .
\label{GammaApprox}
\end{equation}
Setting for $c_n$ and $A_n$ the equations \eqref{wanier_coeficient_final_a}
and \eqref{expansion_coef} we have
\begin{equation}
\Gamma_j\simeq \sum_{m=0}^{j}
\left(\varepsilon_a\alpha_m-\gamma_m\right)
\left(-\alpha_1\right)^{j-m}
\ .
\label{GammaApproxFinal}
\end{equation}
Expressing $\gamma_n$ and $\alpha_n$
by means of the equation \eqref{gamma_alpha_specif 1} we further obtain
\begin{eqnarray}
\Gamma_j
&\simeq&
\left(\varepsilon_a-\gamma_0\right)
\left(-\alpha_1\right)^{j}
\nonumber
\\
&+&
\sum_{m=1}^{j}
\left(\varepsilon_a\alpha_1-\gamma_1\right) e^{-(m-1)B a}
\left(-\alpha_1\right)^{j-m}
\ .
\nonumber
\\
\label{GammaApproxFinal}
\end{eqnarray}
If $\alpha_1 \gg \exp(-Ba)$, we can skip all  terms with $m\geq2$. Thus
\begin{equation}
{\it \Gamma}_j
\simeq
\left(-\alpha_1\right)^{j-1}
\left[
-\gamma_1+\gamma_0\alpha_1) \right]
\simeq
-\left(-\alpha_1\right)^{j-1}\ \gamma_1
\ .
\label{G_by_gamma_fin2A}
\end{equation}


\begin{thebibliography}{00}


\bibitem{Imry-book}
Y.~Imry, \emph{Introduction to Mesoscopic Physics} (Oxford University Press, Oxford, UK, 2002).

\bibitem{Byers}
N.~Byers and C.N. Yang, Phys. Rev. Lett. \textbf{7}, 46 (1961).

\bibitem{Bloch}
F.~Bloch, Phys. Rev. \textbf{137}, A787 (1965); \textbf{166},415 (1968).

\bibitem{Buttiker} M.~B\"{u}ttiker, Y.~Imry, and R.~Landauer, Phys. Lett.~A~{\bf 96},~365~(1983).

\bibitem{Levy} L.~P.~Lévy,~G.~Dolan,~J.~Dunsmuir, and H.~Bouchiat,~Phys.~Rev.~Lett.~{\bf 64}, 2074~(1990).


\bibitem{Chandrasekhar} V.~Chandrasekhar,~R.~A.~Webb,~M.~J.~Brady,~M.~B.~Ketchen, W. J. Gallagher, and A.~Kleinsasser,
    Phys. Rev. Lett.~{\bf 67},~3578~(1991).

    \bibitem{Jariwala} E. M. Q. Jariwala, P. Mohanty, M. B. Ketchen, and R.~A.~Webb,
    ~Phys.~Rev.~Lett.~{\bf 86},~1594~(2001).



\bibitem{Bluhm} H.~Bluhm, N.~C.~Koschnick,~J.~A.~Bert, M.~E.~Huber, and K.~A.~Moler,~Phys.~Rev.~Lett. {\bf 102}, 136802 (2009).

  \bibitem{Bleszynski} A. C. Bleszynski-Jayich, W. E. Shanks, B. Peaudecerf, E. Ginossar,
  F. von Oppen, L. Glazman, and J. G. E. Harris, Science~{\bf 326}, 272 (2009).

  \bibitem{Mailly} D.~Mailly,~C.~Chapelier,~A.~Benoit,~Phys.~Rev.~Lett. {\bf 70}, 2020 (1993).
  \bibitem{Rabaud} W.~Rabaud,~L.~Saminadayar,~D.~Mailly,~K.~Hasselbach,~A.~Beno\^{i}t, B.~Etienne,
    ~Phys.~Rev.~Lett.~{\bf 86},~3124~(2001).

\bibitem{Eckern}
U. Eckern, and P. Schwab, J. Low Temp. Phys. \textbf{126}, 1291 (2002).

\bibitem{Saminadayar} L.~Saminadayar, C.~Bauerle, and D.~Mailly, Encycl. Nanosci. Nanotech.~{\bf 3}, 267 (2004).

\bibitem{Feilhauer2}
J.~Feilhauer and M.~Mo\v{s}ko, Phys.~Rev.~B~{\bf 84}, 085454 (2011).

\bibitem{Mosko} M.~Mo\v{s}ko, unpublished (2006).

\bibitem{Cheung} H.-F.~Cheung,~Y.~Gefen, E.~K.~Riedel, W.-H.~Shih, Phys.~Rev.~B~{\bf 37}, 6050 (1988).

\bibitem{Kohn1}
W. Kohn,  Phys.~Rev.~{\bf
133},~A171~(1964).

\bibitem{Moskova} A.~Mo\v{s}kov\'{a}, M.~Mo\v{s}ko, and A.~Gendiar, Physica {\bf E}, 1991 (2008).

\bibitem{Nemeth-1} R. N\'{e}meth, M. Mo\v{s}ko, Physica E \textbf{40}, 1498 (2008).

\bibitem{Pinsky} M. A. Pinsky, Introduction to Fourier Analysis and Wavelets, (American Mathematical Society, 2002).

\bibitem{Ashcroft}
N.~W. Ashcroft, and N.~D. Mermin, \emph{Solid state physics}, Sounders College Publishing, Ed. D.
G. Crane, Cornell University, USA 1976.


\bibitem{Kohn2}
W. Kohn, Phys.~Rev.~B~{\bf
7},~4388~(1973).

\bibitem{Andreoni}
W. Andreoni, Phys.~Rev.~B~{\bf
14},~4247~(1976).


\bibitem{comment_Wannierfucntions}
Precisely, equation \eqref{wanier_by_atomic approx approx} shows the decay
 ${\mid n \mid}^{-1/2} {\mid \alpha_1 \mid}^{\mid n \mid}$.
A purely exponential decay was predicted by W. Kohn,  Phys.~Rev.~{\bf 115}, 809 (1959).
A very sophisticated estimates by L. He and D. Vanderbilt, Phys. Rev. Lett.~{\bf 86},~5341~(2001), and by A. Bruno-Alfonso and
D. R. Nacbar, Phys.~Rev.~B~{\bf 75}, 115428 (2007), predict (in our notations) the decay
${\mid n \mid}^{-3/4} {\mid \alpha_1 \mid}^{\mid n \mid}$.
For our purposes, exact nature of the power law prefactor is of minor importance.


\bibitem{Bleszynski2} A. C. Bleszynski-Jayich, W. E. Shanks, and J. G. E. Harris,
  Appl.~Phys.~Lett~{\bf 92}, 013123 (2008).

  \bibitem{Bleszynski3} A. C. Bleszynski-Jayich, W. E. Shanks, R. Ilic, and J. G. E. Harris,
  J.~Vac.~Sci.Technol~~{\bf 26}, 1412 (2008).



\bibitem{comment1}
 In fact, the details of the $\Gamma_n$ dependence of the lowest band are not essential
 for the resulting persistent current since
the dominant contribution to the current is due to the light-hole band (see the text). We note only
for completeness that in the GaAs and
 InAs  the $\Gamma_N$ dependence of the lowest band exhibits the same sign for all $N$ except
  for a few isolated values of $N$  where the sign is opposite. We plot the whole $\Gamma_N$ curve of the lowest band
  with a fixed sign for simplicity. Due to this simplification the $\Gamma_N$
curve exhibits a few local minima: the $\Gamma_N$ values at these minima in fact do not have the
same sign as the rest of the curve.


\bibitem{Chadi-Cohen} D.J.~Chadi and M.L.~Cohen, Phys.~Status Solidi~B~{\bf 68},
405~(1975).

\bibitem{Loehr}
J. P. Loehr, and D. N. Talwar, Phys. Rev.~B~{\bf 55}, 4353 (1997).


\bibitem{Mosko-PRB}
M.~Mo\v{s}ko, R.~N\'{e}meth, R.~Kr\v{c}m\'{a}r, and M.~Indlekofer, Phys.~Rev.~B~{\bf
79},~245323~(2009).

\bibitem{Nemeth-PRB}
R. N\'{e}meth, M. Mo\v{s}ko, R. Kr\v{c}m\'{a}r, A. Gendiar, K. M. Indlekofer, and L. Mitas,
arXiv:0902.2225 (2009).

\bibitem{Resta}
R. Resta, Eur. Phys. J.~B~{\bf
79},~121~(2011).

\end{thebibliography}
\end{document}